%% file: article.tex
\documentclass{article}

\usepackage{arxiv}

\usepackage[utf8]{inputenc}
\usepackage[T1]{fontenc}
\usepackage[hidelinks]{hyperref}
\usepackage{url}
\usepackage{booktabs}
\usepackage{amsfonts}
\usepackage{nicefrac}
\usepackage{microtype}
\usepackage{graphicx}
\usepackage{xcolor}
\usepackage{titlesec}

\usepackage{amsmath} 
\usepackage[numbers,super]{natbib}
\usepackage{doi}
\usepackage[font=small]{caption}
\usepackage{multirow,booktabs,array} 
\usepackage{csquotes}

\definecolor{replyblue}{rgb}{0,0,0}

\hypersetup{
pdftitle={Shared representations in brains and models reveal a two-route cortical organization during scene perception},
pdfsubject={q-bio.NC},
pdfauthor={Pablo Marcos-Manchón, Lluís Fuentemilla},
pdfkeywords={Visual Cortex, Scene Perception, Representational Geometry, Deep Neural Networks, Cortical Organization},
}

\title{\textcolor{replyblue}{Shared representations in brains and models reveal a two-route cortical organization during scene perception}}

\author{ 
Pablo Marcos-Manchón \textsuperscript{\textnormal{1,2}} \\
\texttt{pmarcos@ub.edu} \And
Lluís Fuentemilla \textsuperscript{\textnormal{1-3}} \\
\texttt{llfuentemilla@ub.edu}
}

\date{
\textsuperscript{1}Department of Cognition, Development and Education Psychology, Faculty of Psychology,\\ University of Barcelona, Spain \\
\textsuperscript{2}Institute of Neurosciences, University of Barcelona, Spain \\
\textsuperscript{3}Bellvitge Institute for Biomedical Research, Spain
}

\renewcommand{\headeright}{}
\renewcommand{\undertitle}{}
\renewcommand{\shorttitle}{}

\begin{document}
\maketitle

\begin{abstract}

\textcolor{replyblue}{
The brain transforms visual inputs into high-dimensional cortical representations that support diverse cognitive and behavioral goals. Characterizing how this information is organized and routed across the human brain is essential for understanding how we process complex visual scenes. Here, we applied representational similarity analysis to 7T fMRI data collected during natural scene viewing. We quantified representational geometry shared across individuals and compared it to hierarchical features from vision and language neural networks. This analysis revealed two distinct processing routes: a ventromedial pathway specialized for scene layout and environmental context, and a lateral occipitotemporal pathway selective for animate content. Vision models aligned with shared structure in both routes, whereas language models corresponded primarily with the lateral pathway. These findings refine classical visual-stream models by characterizing scene perception as a distributed cortical network with separable representational routes for context and animate content.}
\end{abstract}


\section{Introduction}\label{introduction}

\textcolor{replyblue}{Sensory inputs arrive at the brain as physical signals, but perception depends on how those signals are converted into structured patterns of cortical activity \cite{GrillSpector2004,DiCarlo2012}. In vision, this conversion spans multiple stages, from early visual cortex to downstream regions whose activity patterns encode objects, scenes, and higher-order interactions \cite{DiCarlo2012,goodale1992separate,EpsteinKanwisher1998,Rolls2024}. Characterizing how these representations are transformed across the cortex is central to understanding how the brain performs this computation: identifying which regions encode specific aspects of the sensory input and mapping the networks that route this information between them.}

\textcolor{replyblue}{Naturalistic neuroimaging provides a way to approach this problem. When different individuals view the same stimuli, activity patterns show reliable synchrony across widespread regions \cite{Hasson2004,Baldassano2018,Chen2017}. This inter-subject synchrony is interpreted as evidence that, despite individual variability, a substantial stimulus-locked component is encoded similarly across brains. This shared component can be used to localize regions that reliably encode stimulus-driven information across individuals \cite{Nastase2019,Finn2020}. Representational geometry formalizes this idea by describing each region in terms of pairwise similarities among stimulus-evoked responses, enabling comparisons across individuals and regions, and relating representational variance to stimulus properties that organize it \cite{Kriegeskorte2008,Haxby2014,Sucholutsky2025}.}

\textcolor{replyblue}{Representational geometry also enables direct comparisons between brains and deep neural networks (DNNs). Across vision and language models, internal activations can resemble cortical representations when both systems process the same stimuli \cite{Schrimpf2020,Caucheteux2022,Simony2024}, reflecting shared organizing dimensions that group stimuli according to similar coarse principles \cite{chen2024}. In the visual domain, vision models show graded correspondence with cortical organization, with deeper layers capturing increasingly complex shape- and object-based features \cite{Khaligh2014,Yamins2014,Cichy2016,Konkle2022}. Complementarily, language models capture semantic regularities that abstract away from visual details, with recent work identifying correspondence in higher-level ventral and lateral regions where perceptual information interfaces with conceptual knowledge \cite{Popham2021,Wang2023}. Comparing model-derived representations to brain responses therefore provides a compact way to situate regions along an ordered computational hierarchy and to distinguish visually grounded features from more abstract semantic representations \cite{Cichy2019,Doerig2023}.}

\begin{figure}[!hb]
    \centering
    \includegraphics[width=\textwidth]{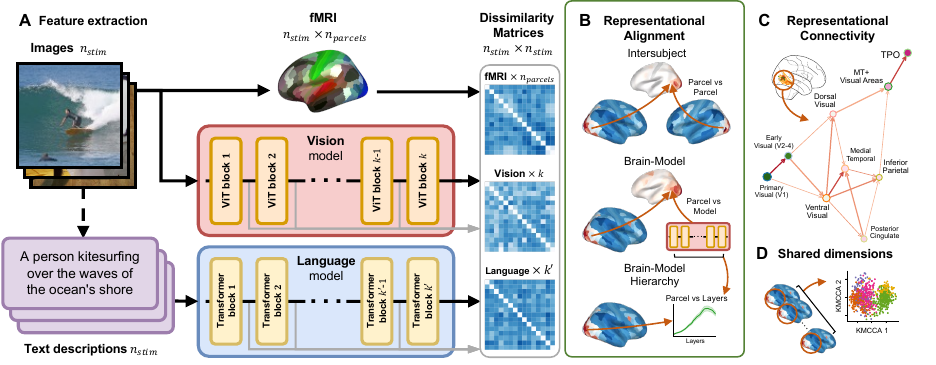}
    \caption{\textbf{Overview of the analysis pipeline.}
\textbf{(A) Feature extraction.} For each image stimulus, we extracted corresponding representations from brain activity and deep neural networks. Single-trial fMRI responses were aggregated within cortical parcels to create vector representations of the brain's response. Concurrently, layer-wise activations were extracted from pre-trained vision and language models to obtain representations across the full model hierarchies. Both brain and model vectors were used to compute representational dissimilarity matrices (RDMs). Example input image adapted from Wikimedia (Bengt Nyman, CC BY 2.0).
\textbf{(B) Representational alignment.} Representational Similarity Analysis (RSA) was used in two ways: (i) inter-subject RSA (IS-RSA), correlating parcel-wise RDMs across participants to estimate shared representational geometry; and (ii) brain–model RSA, correlating parcel RDMs with model-layer RDMs to quantify brain–model and layer-wise alignment profiles.
\textbf{(C) Representational connectivity.} IS-RSA between pairs of parcels was used to construct a cortical network based on how similarly regions encode the stimulus set. The directionality of information flow was inferred from the peak model-layer alignment established in (B).
\textbf{(D) Shared dimensions.} Within the main hubs identified in (C), Kernel Multi-view Canonical Correlation Analysis (KMCCA) was used to decompose the shared geometry into latent dimensions common across participants and to relate these dimensions to coarse scene features that drive alignment in different parts of the network.
}
\label{fig:methodology}
\end{figure}

\textcolor{replyblue}{Here, we investigated how information about visual scenes is encoded and routed across cortex by analyzing representational geometry shared among individuals and DNNs. We introduced a unified framework that links shared representational structure across individuals, its correspondence to vision and language models, and the network topology implied by representational connectivity, while using dimensionality decomposition to relate these patterns to coarse stimulus properties (Fig.~\ref{fig:methodology}). Integrating these complementary approaches, which are often used in isolation, allowed us to move beyond localizing regions with shared information to characterizing how these regions organize as a distributed network and identifying the specific stimulus features that drive this organization.}

\textcolor{replyblue}{
We applied this framework to the Natural Scenes Dataset (NSD) \cite{Allen2022}, in which participants performed a memory recognition task while viewing thousands of natural scenes spanning diverse environments, objects, and social interactions. We recovered a cortex-wide organization with two dissociable processing routes, each specialized to represent distinct types of stimulus information. The medial-ventral stream was specialized for scene layout and environmental context, while the lateral stream was selectively tuned to animate and social content. DNNs captured the shared geometry of both routes, though with clear modality-specific differences: vision models aligned with both pathways, whereas language models corresponded primarily with the lateral stream. These findings are consistent with recent refinements of classical visual-system accounts \cite{goodale1992separate,Kravitz2013}, including proposals of a ventromedial route for scene context and a lateral ``third visual pathway'' for social perception \cite{Rolls2024Review,Pitcher2021}.We then compared these results using two independent datasets (BOLD5000 \cite{Chang2019} and THINGS-fMRI \cite{Hebart2023}) and found that, although the same cortical areas consistently emerged in the alignment analysis, replicating the full two-stream organization required rich scene and social content.}

\section{Results}
\label{results}

\subsection{Inter-subject shared geometry during scene viewing}
\label{sec:intersubject}

\textcolor{replyblue}{To localize cortical regions whose response patterns reflect reliable stimulus information in a shared format across people, we used inter-subject representational similarity analysis (IS-RSA) \cite{Finn2020,Kriegeskorte2008}. We applied this approach to the Natural Scenes Dataset (NSD) \cite{Allen2022}, which provides high-resolution 7T fMRI responses from eight participants performing an inter-session long-term recognition memory task in which they detected repeated images while viewing thousands of natural scenes. Our goal was to measure the stimulus-driven representational geometry related to visual content.}

\textcolor{replyblue}{To isolate this geometry, we parcellated the cortical surface into 180 regions per hemisphere using the Human Connectome Project multimodal parcellation (HCP-MMP) atlas \cite{Glasser2016} and extracted the provided single-trial responses for each image. For each subject and parcel, we constructed representational dissimilarity matrices (RDMs) from pairwise distances between multivoxel patterns. We then quantified inter-subject alignment by correlating parcel-wise RDMs across participants, using a shifted-repetition scheme to reduce mnemonic and session-related confounds (Supplementary Note~\ref{subsection:repetition-matching}). This yielded a parcel-wise map of shared representational geometry.}

\begin{figure}[p]
    \centering
    \includegraphics[width=0.98\textwidth]{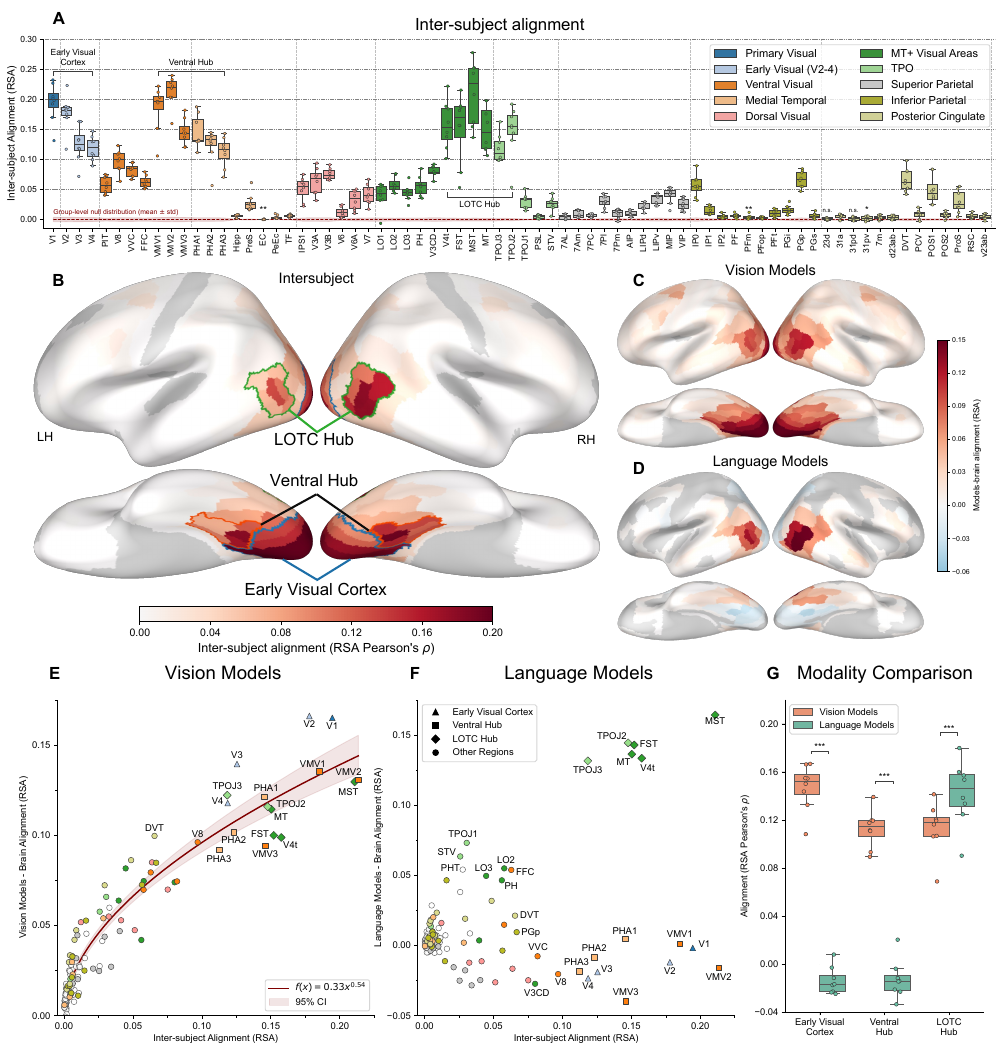}
\caption{\textbf{Inter-subject and model–brain alignment across cortex.}
\textbf{(A)} Inter-subject representational alignment (IS-RSA; Pearson’s $r$ between parcel-wise RDMs across participants) for parcels drawn from the macro-anatomical clusters with the highest mean IS-RSA (color legend). Boxplots show the distribution across subjects in the NSD sample ($N=8$, symmetric HCP–MMP atlas). Red lines indicate the parcel-wise null distribution (mean $\pm$ s.d.; 10{,}000 permutations). Unless otherwise annotated in the plot, parcels are significant at *** ($p<0.001$; two-tailed, FDR-corrected); additional annotations indicate ** ($p<0.01$), * ($p<0.05$), or n.s.\ ($p\ge 0.05$). Peak alignment occurs in early visual cortex (V1–V4), a ventral hub (VMV1–3, PHA1–3), and a lateral occipitotemporal (LOTC) hub (V4t, MT, MST, FST, TPOJ2–3). Full-atlas results are shown in Supplementary Fig.~\ref{fig:extended_spatial_distribution}.
\textbf{(B–D)} Cortical surface maps of alignment, averaged across subjects (and across models within each modality): \textbf{(B)} inter-subject IS-RSA; \textbf{(C)} vision–model RSA (maximum across layers, averaged across vision models); \textbf{(D)} language–model RSA (maximum across layers, averaged across language models).
\textbf{(E–F)} Parcel-wise relationship between IS-RSA and model–brain alignment for \textbf{(E)} vision models and \textbf{(F)} language models. Points are colored by macro-anatomical group (as in panel \textbf{A}); other parcels are shown in white. Vision parcels follow an approximate power-law fit ($R^{2}=0.94$, shaded band: 95\% bootstrap CI), whereas language-model alignment clusters near zero or negative except for parcels in and around the LOTC hub.
\textbf{(G)} Modality comparison within the three hubs. Boxplots show hub alignment for vision and language models in early visual cortex, the ventral hub, and the LOTC hub (averaged across models). Paired two-tailed $t$-tests ($t(7)$, FDR-corrected) indicate stronger vision-model alignment in early visual and ventral hubs, and stronger language-model alignment in the LOTC hub.
}  
\label{fig:spatial}
\end{figure}

\textcolor{replyblue}{IS-RSA captured significant alignment across occipital cortex and extended into higher-order association cortex (Fig.~\hyperref[fig:spatial]{\ref*{fig:spatial}a}; Supplementary Figs.~\ref{fig:extended_spatial_distribution}-\ref{fig:cortical_surfaces}). Alignment was strongest in early visual cortex (V1-V4) and extended into ventral occipitotemporal and dorsal occipitoparietal regions, consistent with hierarchical accounts of visual organization \cite{DiCarlo2012, Kravitz2013, Felleman1991}. We also observed significant, albeit weaker, alignment in prefrontal regions, which may reflect shared contributions of visual attention and task-dependent semantic processing\cite{Freedman2008,Weiner2011,Bugatus2017}. Within this map, three anatomically clustered sets of parcels showed especially strong and consistent alignment across participants (Fig.~\hyperref[fig:spatial]{\ref*{fig:spatial}b}).}

\textcolor{replyblue}{The first cluster corresponds to the early visual cortex (V1-V4), consistent with its role in encoding low-level visual features \cite{Wandell2007}. The second comprised a ventral hub in ventromedial temporal cortex (VMV1-3, PHA1-3), overlapping scene-selective regions such as the parahippocampal place area (PPA) and adjacent cortex implicated in encoding scene layout and place memory \cite{EpsteinKanwisher1998,Rolls2024,Park2011}. The third comprised a lateral occipitotemporal cortex (LOTC) hub including the motion-selective MT+ complex and posterior temporoparietal junction (MT, MST, FST, V4t, TPOJ2-3), near regions responsive to bodies and biological motion that have been implicated in social perception \cite{Pitcher2021, Weiner2011, Beauchamp2004}. In the following sections, we use these three hubs (early visual, ventral, and LOTC) as reference points for subsequent analyses.}

\textcolor{replyblue}{This spatial distribution is consistent with prior NSD work reporting variability in signal reliability and decodability across cortex \cite{Allen2022,Takagi2023,Conwell2024,Doerig2025}. Across parcels, higher IS-RSA values were associated with higher signal reliability (Supplementary Note~\ref{subsection:resampling-reliability}), and a within-subject RSA analysis comparing different repetitions within each participant produced a similar cortical pattern (Supplementary Fig.~\ref{fig:within-subject}). Although alignment strength showed some hemispheric asymmetries (Fig.~\hyperref[fig:spatial]{\ref*{fig:spatial}b}), representational profiles of left-right homologous parcels were highly correlated, consistent with a largely bilateral organization (Supplementary Note~\ref{subsection:asymmetry}). Together, these results indicate that IS-RSA isolates a distributed set of regions whose scene-evoked representational geometry is shared across individuals, providing a benchmark for model comparisons described in the next sections.}

\subsection{Model–brain alignment reveals modality-specific shared geometry}
\label{subsec:model-brain}

\textcolor{replyblue}{Although IS-RSA localized parcels whose representational geometry was shared across observers, it did not by itself specify which stimulus information underlay that shared structure. To distinguish visually grounded from more abstract features, we compared cortical RDMs with RDMs from pretrained vision and language models. Vision models provide hierarchies learned directly from pixels \cite{Zeiler2014,Kriegeskorte2015,Yamins2016}, whereas large language models (LLMs) provide a reference for text-derived structure. By contrasting brain alignment with these two model classes, we characterized whether a parcel's shared geometry was better captured by visually grounded features or by higher-level language representations.}

\textcolor{replyblue}{We analyzed 37 pretrained models based on Vision Transformer (ViT) \cite{dosovitskiy2021an} and Transformer \cite{Vaswani2017} architectures, spanning supervised, self-supervised, and contrastive objectives (Table~\ref{tab:foundation_models}). For each model, we extracted layer-wise activations for each NSD image (vision models) or its caption (language models), constructed layer-specific RDMs, and quantified model-brain alignment by correlating each layer RDM with each parcel RDM. For each model, parcel-wise alignment was summarized as the maximum RSA across layers. Fig.~\ref{fig:spatial}c-d show per-parcel maps averaged across subjects and, within each modality, across models (significance was assessed with a permutation-based null distribution).}

\textcolor{replyblue}{Vision models produced an alignment map that closely matched the IS-RSA spatial pattern (Fig.~\hyperref[fig:spatial]{\ref*{fig:spatial}c} vs.\ Fig.~\hyperref[fig:spatial]{\ref*{fig:spatial}b}). Parcels with high IS-RSA also showed strong vision-model alignment, with peaks in the Early Visual, Ventral, and LOTC hubs and weaker but significant effects across additional visual and prefrontal parcels. This overlap indicates that image-trained feature spaces capture a substantial component of the stimulus-structured geometry that is shared across individuals during scene viewing, consistent with prior NSD results and ventral-stream model-brain alignment work \cite{Khaligh2014,Yamins2014,Cichy2016,Wang2023, Allen2022,Takagi2023}. Across parcels, IS-RSA and vision-model alignment were strongly related by approximate power-law scaling ($R^2=0.94$; Fig.~\hyperref[fig:spatial]{\ref*{fig:spatial}e}), consistent with both measures being largely driven by a common stimulus-locked component of scene organization (see Supplementary Note~\ref{section:noise-ceiling} for analysis of scaling and reliability).}

\textcolor{replyblue}{In contrast, autoregressive LLMs trained on next-token prediction showed alignment concentrated in LOTC and adjacent temporal cortex, and more weakly in anterior areas of visual cortex (Fig.\hyperref[fig:spatial]{\ref*{fig:spatial}d,f}), regions implicated in biological motion, social perception, and multimodal processing \cite{Popham2021, Beauchamp2004,Pitcher2021,Weiner2011}. Alignment in these areas is consistent with prior reports of language-model correspondence in lateral/anterior temporal cortex and medial frontal regions in naturalistic paradigms associated with semantic processing \cite{Popham2021,Wang2023,Doerig2025}. Within the LOTC hub, language-model alignment exceeded vision-model alignment (Fig.\hyperref[fig:spatial]{\ref*{fig:spatial}g}), indicating that the shared geometry expressed in this hub is better captured by caption-derived structure than by purely visual model features.}

\textcolor{replyblue}{Conversely, early and ventral visual regions showed near-zero or negative RSA with these LLM embeddings (Fig.~\hyperref[fig:spatial]{\ref*{fig:spatial}d}; Supplementary Fig.~\ref{fig:cortical_surfaces}). Negative RSA indicates a systematic mismatch in geometry: stimulus features that separate responses in these parcels, mainly dominated by visual characteristics, are not emphasized by the language-model representations. This lack of alignment contrasts with studies finding ventral correspondence using language models optimized for sentence similarity or language-vision contrastive objectives \cite{Wang2023,Doerig2025}. Unlike the next-token prediction models used here, similarity-based models organize embeddings in their output layer by semantic content--grouping descriptions of similar scenes--which results in geometries that overlap more substantially with high-level visual representations.}

\begin{figure}[!ht]
    \centering
    \includegraphics[width=\textwidth]{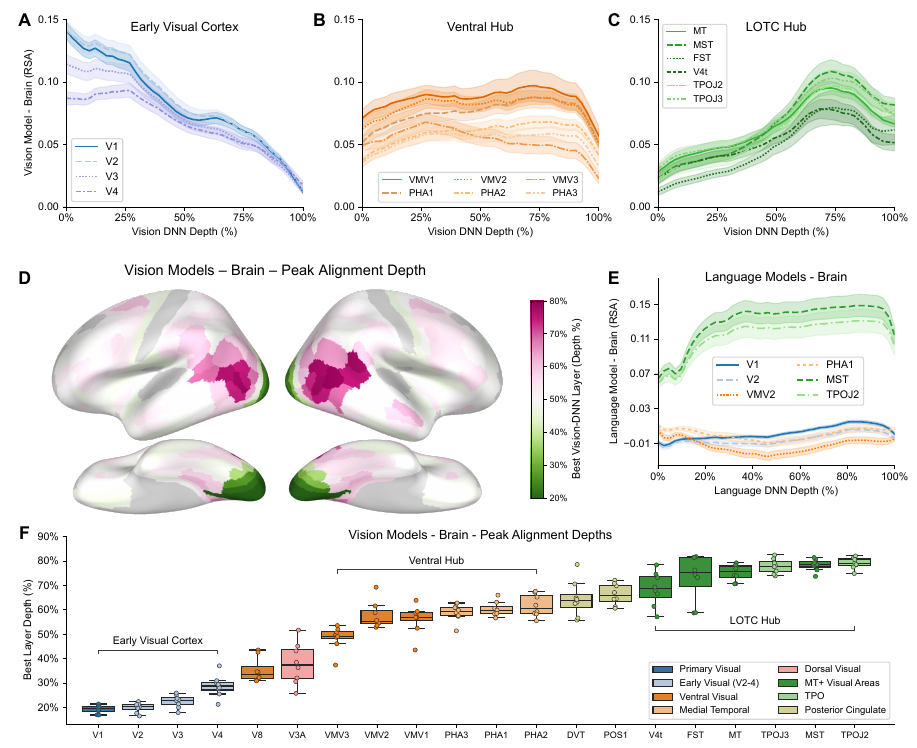}
    \caption{\textbf{Hierarchical correspondence between model layers and cortical regions.}
\textbf{(A–C)} Layer-wise alignment (RSA) between vision models and representative parcels in the three hubs, averaged across all vision models. Panels show early visual cortex (\textbf{A}), the Ventral hub (\textbf{B}), and the LOTC hub (\textbf{C}). The $x$-axis expresses depth as a percentage of the total number of layers in each vision model. Curves show the mean alignment across participants ($N=8$); shaded bands indicate the standard error of the mean (SEM). Early visual parcels peak in shallow layers, Ventral parcels show distributed alignment across the hierarchy, and LOTC parcels peak in the deepest layers.
\textbf{(D)} Cortical surface map of the vision-model layer that yields the highest alignment for each parcel (averaged across vision models). Colors encode normalized peak-layer depth (0\% = shallowest layer, 100\% = deepest layer), revealing a posterior–anterior gradient from shallow (green) to deep (purple) alignment.
\textbf{(E)} Layer-wise alignment between language models and representative parcels from the three hubs (mean across language models). Alignment values are negligible in early visual and Ventral parcels and are restricted to LOTC, where curves rise quickly and then form a plateau rather than a smooth hierarchical progression.
\textbf{(F)} Distribution across participants ($N=8$) of peak alignment depths for the 20 parcels with the highest overall vision–model alignment. For each subject and parcel, peak depth is defined as the mean depth (across vision models) of the layer at which RSA is maximal. Boxplots are ordered by median peak depth and colored by macro-anatomical cluster.}
    \label{fig:hierarchy}
\end{figure}

\subsection{Hierarchical correspondence between models and cortex}
\label{section:hierarchy}

\textcolor{replyblue}{To relate cortical parcels to the ordered representational stages learned by DNNs, we examined how model-brain alignment varied across model depth. Prior work has shown that, for vision models, shallow layers tend to align with early visual cortex and deeper layers tend to align with higher-level ventral and downstream regions \cite{Schrimpf2020,Khaligh2014,Yamins2014,Cichy2016,Guchu2015}. Here, we assessed this depth dependence during natural scene viewing in NSD by analyzing layer-wise RSA curves, focusing on the three hubs where inter-subject shared geometry was strongest (Fig.~\ref{fig:hierarchy}; full parcel-wise curves and model-family profiles in Supplementary Figs.~\ref{fig:extneded_roi_curves}-\ref{fig:extneded_family_curves}).}

\textcolor{replyblue}{The three hubs showed distinct depth profiles for vision models (Fig.~\hyperref[fig:hierarchy]{\ref*{fig:hierarchy}a-c}). Early visual parcels peaked in the shallowest layers and then decreased (Fig.~\hyperref[fig:hierarchy]{\ref*{fig:hierarchy}a}), consistent with correspondence to lower-level visual features. Ventral hub parcels showed broad alignment spanning earlier and later layers (Fig.~\hyperref[fig:hierarchy]{\ref*{fig:hierarchy}b}), indicating that their shared geometry relates to features distributed across multiple stages of the vision-model hierarchy rather than a single best-matching depth. LOTC parcels increased more steadily with depth and reached their maxima in deeper layers (Fig.~\hyperref[fig:hierarchy]{\ref*{fig:hierarchy}c}), indicating that the stimulus organization expressed in LOTC is best matched by later-stage vision-model features.}

\textcolor{replyblue}{To summarize this organization across cortex, we identified for each parcel the vision-model layer with maximal alignment and projected this peak depth onto the cortical surface. The resulting map formed a posterior-to-anterior gradient in which early visual cortex was best explained by shallower layers, while more anterior occipitotemporal, STS-adjacent, and prefrontal regions were best explained by deeper layers (Fig.~\hyperref[fig:hierarchy]{\ref*{fig:hierarchy}d}). This gradient is consistent with hierarchical accounts of visual processing \cite{Felleman1991,DiCarlo2012,Kravitz2013,Rolls2024} and with prior work showing that model depth tracks representational complexity across the human ventral pathway \cite{Guchu2015,Yamins2014,Yamins2016,Khaligh2014,Cichy2016,Schrimpf2020}. We use peak depth as a descriptive index that orders parcels along the model hierarchy (Fig.~\hyperref[fig:hierarchy]{\ref*{fig:hierarchy}f}).}

\textcolor{replyblue}{Language models showed a different pattern. Alignment was concentrated in the LOTC hub, and layer-wise profiles did not show the early-to-late progression observed for vision models. Instead, alignment was present at the first layer, increased, and then approached a plateau. Across the language models we evaluated, first-layer alignment strongly predicted the plateau magnitude (Pearson's $r=0.97$) and was related to the tokenizer representation (Pearson's $r=0.78$; Supplementary Note~\ref{subsection:tokenizer}). This pattern is consistent with evidence that, in visual contexts, brain-language correspondence can be driven by relatively coarse semantic content, closer to a bag-of-concepts organization than to progressively composed syntax \cite{yuksekgonul2023when}. Interpreting this contrast depends on a linking assumption about inputs: vision models and cortex are compared while transforming pixel-level images, whereas language models operate on tokenized captions that already provide an abstracted description. Under this input mismatch, language-model alignment can emerge in regions that express higher-level scene organization, such as LOTC, without producing a depth-dependent gradient across the visual hierarchy.}

\begin{figure}[ht]
    \centering
    \includegraphics[width=\textwidth]{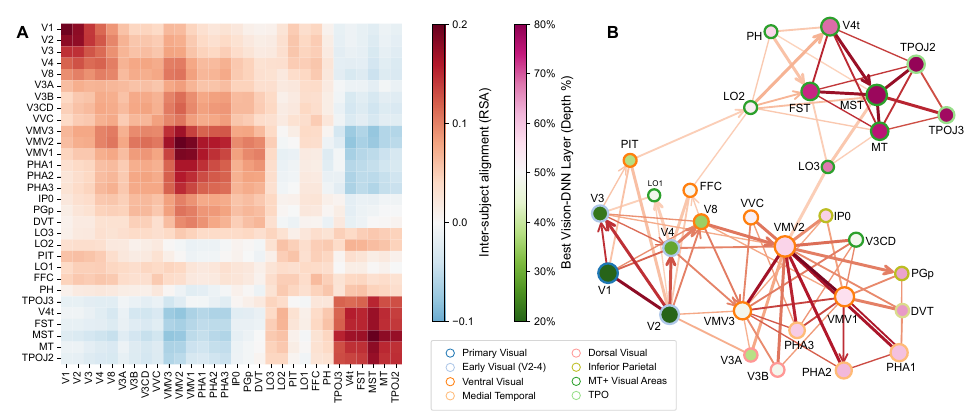}
     \caption{\textbf{Inter-subject representational connectivity network.}
\textbf{(A) Inter-subject connectivity matrix.} Parcel-wise representational connectivity (RSA; Pearson’s $r$) for the 30 parcels with the highest mean IS-RSA ($N=8$). Each cell shows the mean RSA between the RDMs of two parcels, computed across all cross-subject pairs. The matrix shows three clusters corresponding to early visual cortex (V1–V4), the Ventral hub (VMV1–3, PHA1–3), and the LOTC hub (V4t, MT, MST, FST, TPOJ2–3). Warm colors indicate higher representational similarity; cool colors indicate dissimilarity.
\textbf{(B) Directed connectivity graph.} The same parcels are shown as nodes. Node color encodes each parcel’s peak alignment depth with vision models (heuristic proxy for hierarchical position), node size indicates inter-subject alignment strength, and node border color indicates macro-anatomical group membership. Edges correspond to the union of three minimum spanning trees; edge color reflects the RSA value in panel \textbf{A}, and edge width indicates the spanning-tree iteration (first to third). Arrow direction is assigned from shallower to deeper peak-depth parcels when the depth difference is significant across subjects (paired $t$-test, $t(7)$, FDR-corrected $p<0.05$). The graph highlights two dominant routes from early visual cortex: a medial-ventral route toward the Ventral hub and a lateral route toward the LOTC hub.
}
    \label{fig:connectivity}
\end{figure}

\subsection{A representational network linking ventromedial and lateral hubs}
\label{section:connectivity}

\textcolor{replyblue}{To characterize how regions with shared scene-evoked information relate to each other, we repurposed IS-RSA as a measure of representational connectivity. This defines a task- and stimulus-dependent network based on similarity in how stimuli are organized across regions, rather than on anatomical proximity or raw BOLD covariance. For each pair of parcels, we correlated their RDMs across different individuals, yielding a parcel-by-parcel connectivity matrix that quantifies how similarly distinct regions organize the stimulus set across observers (Fig.~\ref{fig:connectivity}a). This approach emphasizes the informational content of the connectivity by grouping regions that share a common representational format, complementing univariate and conventional functional-connectivity analyses \cite{Pillet2019,Huang2024}.}

\textcolor{replyblue}{The connectivity matrix showed three clusters aligned with the Early Visual, Ventral, and LOTC hubs (Fig.~\hyperref[fig:connectivity]{\ref*{fig:connectivity}a}). Early visual and ventral parcels formed a broad posterior cluster that also included the occipital and parietal cortex, indicating similar stimulus organization across these regions. LOTC parcels formed a second cluster with low or negative similarity to the posterior cluster. The strongest links between these systems passed through intermediate parcels in lateral occipital and ventral temporal cortex (LO1-3, FFC, PH). Because IS-RSA reflects both shared structure and measurement noise, local reliability differences can affect edge strength. However, the same topology was recovered with noise-ceiling normalization (Supplementary Fig.~\ref{fig:noise-ceil-connectivity}) and in a within-subject analysis (Supplementary Fig.~\ref{fig:within-subject}).}

\textcolor{replyblue}{To visualize the strongest routes, we extracted a sparse backbone graph by iteratively applying a minimum spanning tree algorithm to the parcel-wise connectivity matrix. We then oriented edges using each parcel's peak alignment depth from the vision-model analysis (Fig.~\hyperref[fig:hierarchy]{\ref*{fig:hierarchy}d,f}) as a descriptive proxy for representational level, from shallower-layer peaks toward deeper-layer peaks. This ordering is not anatomical and does not imply causal direction. In the resulting graph, early visual cortex bifurcated into two routes (Fig.~\hyperref[fig:connectivity]{\ref*{fig:connectivity}b}): a medial-ventral path progressing through ventral occipitotemporal cortex into the Ventral hub, and a lateral path progressing through lateral occipital cortex into the LOTC hub. Extending the analysis to the whole brain (Supplementary Fig.~\ref{fig:whole_connectivity}) showed both routes continuing into association cortex, consistent with these representational formats interfacing with downstream systems implicated in memory, action, and control\cite{MillerCohen2001, Ranganath2012}.}

\textcolor{replyblue}{This two-route organization does not map cleanly onto the classical ``ventral what / dorsal where'' framing of visual processing \cite{goodale1992separate,Felleman1991,Kravitz2013,VaziriPashkam2024TwoWhat}. The medial-ventral branch targeted parahippocampal and ventromedial visual parcels (including VMV and parahippocampal scene regions) that are consistently linked to scene layout and environmental context \cite{EpsteinKanwisher1998,Baldassano2016,Epstein2019}, and it matches proposals of a ventromedial, scene-based pathway that provides scene input to hippocampal memory systems \cite{Rolls2024,Rolls2024Review}. The lateral route, by contrast, encompassed MT+ and LOTC and continued toward posterior STS and temporo-parietal cortex in the whole-brain graph, regions implicated in dynamic cues, biological motion, body perception, and social processing \cite{Beauchamp2004,Pitcher2021,Rolls2024Review,Pitcher2025,Choi2020}. In this representational network, regions typically associated with the classical dorsal stream in intraparietal and superior parietal cortex were less prominent in shared geometry, but they remained part of the broader posterior cluster. This pattern suggests that the LOTC-centered trajectory reflects a distinct lateral pathway whose shared stimulus organization is driven most strongly by biological and socially relevant content, rather than the spatial computations emphasized in traditional ``where/how'' accounts \cite{Pitcher2021}.}

\subsection{Representational dimensions driving shared geometry}

\begin{figure}[t]
    \centering
    \includegraphics[width=\textwidth]{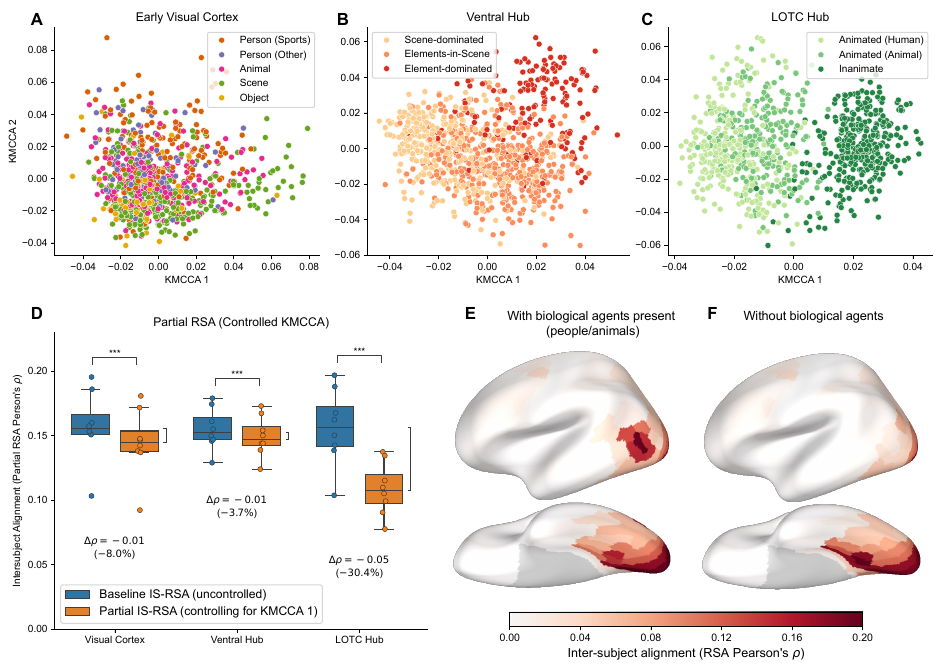}
     \caption{\textbf{KMCCA components and biological-content dependence of LOTC shared geometry.}
\textbf{(A-C)} Stimulus projections onto the first two KMCCA components for each hub, estimated from all eight NSD participants using the shared image set. Early Visual Cortex (\textbf{A}) shows no clear clustering by scene category. The Ventral hub (\textbf{B}) varies along a scene-object continuum. The LOTC hub (\textbf{C}) shows a categorical separation along an animacy-related axis (animate agents vs.\ non-animate scenes/objects).
\textbf{(D)} Effect of removing the first KMCCA component (KMCCA1) using partial RSA. Orange boxplots show partial IS-RSA after regressing out KMCCA1 from the hub RDMs; blue shows the original IS-RSA. Removing this single dimension reduces inter-subject alignment in all hubs, with the largest reduction in LOTC. IS-RSA was computed per parcel and subject, averaged within hub, and then averaged across subjects; paired $t$-tests were Bonferroni-corrected ($t(7)$, $p<0.001$). Annotated $\Delta\rho$ reports the mean absolute change, and percentages denote the relative change normalized by baseline.
\textbf{(E-F)} Inter-subject RSA maps recomputed after splitting the stimulus set into scenes \emph{with} biological agents (\textbf{E}) versus \emph{without} biological agents (\textbf{F}). LOTC alignment is strongly reduced when biological content is absent.}
\label{fig:components}
\end{figure}

\textcolor{replyblue}{The two-route organization of the scene network raises a concrete question: which stimulus information most strongly organizes the shared representational geometry within each hub and along each route. To address this directly from the data, under the present task context, we decomposed shared across-subject structure in the Early Visual, Ventral, and LOTC hubs using Kernel Multi-view Canonical Correlation Analysis (KMCCA) \cite{Hardoon2004MKCCA}. Unlike RSA, which summarizes similarity between two RDMs with a single correlation, KMCCA treats each participant’s hub RDM as a separate view of the same stimulus set and learns components that maximize correlation across views. We applied multi-view CCA ideas \cite{Sui2012} to RDM-derived kernels, yielding a low-dimensional set of components that captures the geometry shared across participants within each hub.}

\textcolor{replyblue}{To visualize this shared geometry, we projected stimuli onto the first two KMCCA components for each hub (Fig.~\ref{fig:components}a-c). In early visual cortex, the projection showed a diffuse distribution with no clear clustering by scene category, consistent with representations dominated by low-level image structure under this stimulus set and task \cite{Doerig2025} (Fig.~\ref{fig:components}a; Supplementary Fig.~\ref{fig:atlas_evc}). In the Ventral hub, the leading axis formed a continuous gradient from panoramic scenes to more object-like images, consistent with ventromedial scene-object organization and scene layout/context representations \cite{EpsteinKanwisher1998,Rolls2024,Rolls2024Review,Baldassano2016,Epstein2019} (Fig.~\ref{fig:components}b; Supplementary Fig.~\ref{fig:atlas_ventral}). In the LOTC hub, the first component separated images containing biological agents (people/animals) from a broader set of non-biological images, including objects and landscapes (Fig.~\ref{fig:components}c; Supplementary Fig.~\ref{fig:atlas_lotc}), consistent with LOTC sensitivity to biological and social content \cite{Pitcher2021,Weiner2011,Beauchamp2004}.}

\textcolor{replyblue}{We then quantified how much this leading axis contributed to each hub’s shared geometry. We regressed out the first KMCCA component and recomputed inter-subject alignment using partial RSA (Fig.\hyperref[fig:components]{\ref*{fig:components}d}). Removing this single axis reduced IS-RSA in all three hubs (all $p<.001$, Bonferroni-corrected), but the reduction was substantially larger in LOTC ($-30.4\%$) than in Early Visual Cortex ($-8.0\%$; $t(7)=-10.78$, $p<.001$) or the Ventral hub ($-3.7\%$; $t(7)=-12.21$, $p<.001$). This pattern indicates that an animacy-related dimension accounts for a large fraction of LOTC’s shared geometry, whereas Early Visual and Ventral hubs distribute shared variance across higher-dimensional structure (Supplementary Fig.\ref{fig:extended_components}).}

\textcolor{replyblue}{Given the dominant agent-related axis in LOTC, we tested whether lateral-route alignment depended on biological content by splitting the stimulus set into scenes with biological agents (people or animals; 65.3\% of images) versus scenes without them, and recomputing IS-RSA within each subset. Alignment in LOTC was strongly reduced for non-biological scenes but remained robust when biological agents were present (Fig.\hyperref[fig:components]{\ref*{fig:components}e–f}). The same split also weakened the lateral branch of the representational network linking early visual cortex to LOTC, reducing the detectability of the two-route backbone when biological content was absent (Supplementary Fig.~\ref{fig:split_analysis}). This pattern indicates that biological-agent content is a major contributor to LOTC shared geometry in this stimulus set and task regime, consistent with proposals that lateral pathways are preferentially organized by socially and biologically relevant scene information.}

\begin{figure}[p]
    \centering
    \includegraphics[width=\textwidth]{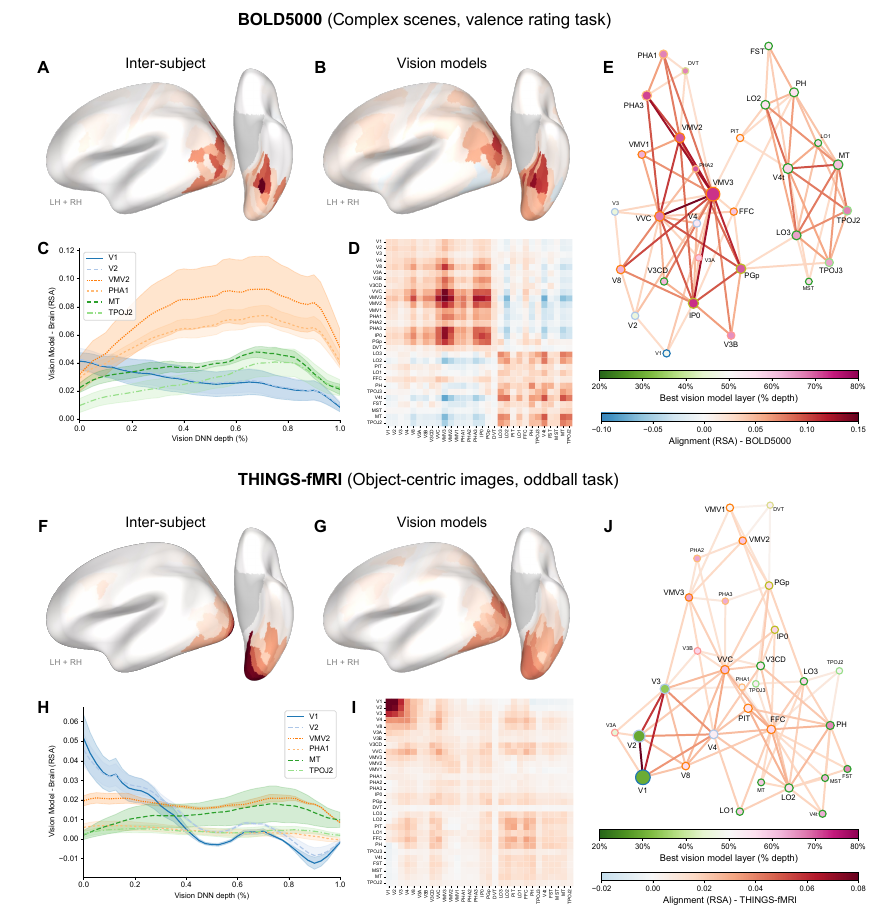}
     \caption{\textbf{Comparison across independent datasets.} Analyses were performed on the BOLD5000 and THINGS-fMRI datasets using the same pipeline as NSD.
\textbf{(A--E) BOLD5000 (complex scenes, valence task; N=4).} Despite lower signal reliability and differences in task, the core network topology persists in this dataset, which shares similar scene statistics with NSD.
\textbf{(A, B)} Inter-subject alignment (\textbf{A}) and vision-model alignment (\textbf{B}) recover the characteristic occipitotemporal footprint, including Ventral and LOTC hubs, though with reduced magnitude and spatial continuity compared to NSD.
\textbf{(C)} Layer-wise alignment profiles (mean $\pm$ SEM across subjects) preserve the hierarchical ordering (Early Visual $<$ Ventral $<$ LOTC) but exhibit flatter curves.
\textbf{(D, E)} Representational connectivity matrix (\textbf{D}) and backbone graph (\textbf{E}) successfully recover the dual-stream topology (based on a $k=3$ minimum spanning tree), segregating the medial--ventral route from the lateral route. In the graph (\textbf{E}), node fill indicates peak model depth and edge color indicates representational connectivity strength.
\textbf{(F--J) THINGS-fMRI (object-centric images, oddball task; N=3).} With minimal scene layout and social content, the network contracts.
\textbf{(F, G)} Inter-subject (\textbf{F}) and vision-model (\textbf{G}) alignment are restricted primarily to early and ventral visual cortex, with no prominent LOTC peak.
\textbf{(H)} Layer-wise profiles (mean $\pm$ SEM) are reduced in magnitude and show flatter profiles, especially in ventral and LOTC hubs.
\textbf{(I, J)} The lateral stream does not emerge as a distinct subsystem in the connectivity matrix (\textbf{I}), and the backbone graph (\textbf{J}) lacks the clear two-route segregation evident in NSD and BOLD5000, consistent with the content-dependence of the lateral pathway.}
    \label{fig:comparison}
\end{figure}

\subsection{Comparison across tasks and stimulus sets}

\textcolor{replyblue}{Our analyses in NSD identified a distributed scene network by quantifying representational geometry shared across individuals and relating it to model feature spaces. Because this geometry is stimulus- and task-dependent, its spatial expression can vary with the regularities present in the stimulus set \cite{Cukur2013,Harel2014}, task demands that reweight the stimulus features that are emphasized \cite{Bugatus2017,Hebart2018}, and measurement reliability, which bounds how well shared structure can be estimated from trial-level data \cite{Nili2014,Diedrichsen2017,Walther2016}. To assess how the inferred organization generalizes across experimental regimes, we applied the same pipeline to two independent datasets: BOLD5000 \cite{Chang2019} and THINGS-fMRI \cite{Hebart2023} (Fig.\ref{fig:comparison}).}

\textcolor{replyblue}{BOLD5000 \cite{Chang2019} uses natural images broadly similar to NSD but differs in task (valence judgment vs.\ memory), stimulus duration (1\,s vs.\ 3\,s), and trial-level reliability under comparable estimation procedures \cite{Allen2022}. In this regime, inter-subject and vision-model alignment remained concentrated in occipital and occipitotemporal cortex, with clear involvement of the ventral and LOTC hubs identified in NSD (Fig.~\ref{fig:comparison}a-b), consistent with both datasets sampling rich scene layout, context, and social content. However, the overall spatial profile differed from NSD in two main respects. Early visual cortex was not detected as a dominant peak, and the LOTC pattern was less spatially contiguous. These differences are likely attributable to reduced signal reliability resulting from the lower field strength (3T) and shorter presentation times. Despite these differences, vision-model alignment preserved a monotonic parcel-wise relationship with IS-RSA, as in NSD (power-law scaling $R^{2}=0.71$). Layer-wise profiles preserved the hub-specific ordering but were flatter than in NSD (Fig.~\ref{fig:comparison}c).}

\textcolor{replyblue}{At the network level, representational connectivity in BOLD5000 still separated a broad posterior cluster spanning early visual, ventral, and adjacent cortex from a lateral occipitotemporal cluster that included LOTC together with nearby lateral occipital and ventral temporal parcels (Fig.~\ref{fig:comparison}d). LOTC was less segregated from neighboring lateral parcels than in NSD, but the backbone visualization still recovered a two-route organization from early visual cortex toward ventral versus lateral targets (Fig.~\ref{fig:comparison}e), consistent with a ventromedial route associated with scene context and a lateral route associated with biological and social features.}

\textcolor{replyblue}{THINGS-fMRI differs from NSD in both stimulus content and task. It uses briefly presented (500\,ms), object-centric images with minimal scene layout and reduced social content, and participants performed an orthogonal oddball task \cite{Hebart2023}. In this regime, inter-subject alignment and vision-model alignment again remained tightly coupled (power-law scaling $R^2=0.80$; Fig.~\ref{fig:comparison}f-g), but the spatial pattern was concentrated in occipital and ventral visual cortex. The NSD ventromedial/parahippocampal and LOTC hubs were not detected as prominent hubs, and LOTC resembled the NSD pattern obtained when restricting analyses to images without biological agents (Fig.~\ref{fig:components}f). Layer-wise profiles retained the same ordering but were reduced in magnitude and less distinctive for the ventral and LOTC hubs (Fig.~\ref{fig:comparison}h). Consistent with this, representational connectivity did not separate an LOTC-centered subsystem (Fig.~\ref{fig:comparison}i), and the backbone graph did not recover the two-route topology observed for NSD and BOLD5000 (Fig.~\ref{fig:comparison}j).}

\textcolor{replyblue}{Across datasets, vision-model alignment continued to track inter-subject alignment, indicating that the same stimulus-locked geometry that is detectable across observers is also present in image-trained feature spaces. However, which hubs and routes emerged as prominent depended on the structure available in the stimulus set and on task/reliability constraints: scene-rich datasets expressed ventromedial/parahippocampal and LOTC involvement more clearly, whereas the object-centric THINGS regime emphasized early and ventral visual cortex and did not express a separable LOTC-centered route.}

\clearpage
\section{Discussion}\label{discussion}

\textcolor{replyblue}{In this study, we aimed to characterize how the human cortex encodes natural scenes by quantifying representational structure shared across individuals and comparing it to hierarchical features in DNNs. We identified a distributed network of cortical regions engaged by scene perception and characterized its major hubs and the stimulus features that organize the processing within it.}

\subsection*{Two representational routes during scene viewing}

\textcolor{replyblue}{We identified this distributed network engaged during scene perception using inter-subject representational connectivity, which captures similarity in how different regions organize the stimulus set.  The resulting topology separated three prominent hubs connected by two dominant routes: an early visual hub, a ventral hub spanning ventromedial and parahippocampal scene regions, and an MT+/LOTC hub. The early visual hub bifurcated into branches toward the ventral and lateral hubs, and the ventral and lateral systems were additionally coupled via intermediate lateral occipital and ventral temporal parcels (Fig.~\ref{fig:connectivity}). This organization does not match the classical ventral ``what'' versus dorsal ``where/how'' partition \cite{goodale1992separate,Kravitz2013}, but is instead consistent with proposals that fractionate visual processing into additional routes, including a ventromedial pathway that emphasizes scene context and a dorsolateral pathway centered on MT+ and STS-adjacent regions that is associated with social/biological information \cite{Rolls2024Review,Pitcher2021,Lima2023}.}

\textcolor{replyblue}{The medial-ventral route aligns with accounts of a ventromedial ``where/scene'' pathway that supports environmental layout and spatial context, providing inputs to hippocampal circuitry\cite{Rolls2024, Rolls2024Review, Rolls2022}. In our data, parcels in this ventral hub showed broad correspondence to vision models across multiple depths (Fig.~\ref{fig:hierarchy}b), indicating that their shared geometry relates to features distributed across multiple stages of the vision-model hierarchy. This pattern held for scenes with and without biological agents (Fig.~\ref{fig:components}e–f). It was reduced in the object-centric THINGS dataset, which contains limited scene layout and context (Fig.~\ref{fig:comparison}e–f), confirming that this route is most detectable when the stimulus set contains rich environmental information. Language-model alignment in these regions showed no correspondence (Fig.~\ref{fig:spatial}d), diverging from findings using similarity-based models \cite{Wang2023,Doerig2025}, indicating that caption-derived similarity structure did not capture the visually grounded scene-layout geometry expressed along the medial-ventral route.}

\textcolor{replyblue}{The lateral route aligns with accounts of an MT+/STS-centered pathway specialized for socially relevant information, including ``third visual pathway'' and dorsolateral subdivision proposals \cite{Rolls2024Review,Pitcher2021,Lima2023}. Here, shared geometry was dominated by biological-agent content. The leading LOTC dimension separated scenes with animate agents from those without, and restricting analyses to images without biological agents reduced LOTC inter-subject alignment and weakened the lateral branch of the network (Fig.~\ref{fig:components}e-f; Supplementary Fig.~\ref{fig:split_analysis}). LOTC parcels aligned most strongly with later stages of vision models (Fig.~\ref{fig:hierarchy}c) and also aligned with language-model representations (Fig.~\ref{fig:spatial}d,f,g), consistent with this route emphasizing socially structured distinctions that are expressed in deep visual features and are also present in text-derived semantic structure.}

\subsection*{Model-brain alignment across the visual hierarchy}

\textcolor{replyblue}{Model-brain RSA related the representational geometry measured in cortex to the ordered progression of features across model depth. For vision models, model-brain correspondence tracked the inter-subject shared geometry across parcels and scaled with IS-RSA with an approximate power-law relationship (Fig.~\ref{fig:spatial}e). This suggests that the geometry shared across observers during scene viewing is also present in image-trained feature spaces, consistent with the view that a reduced set of coarse dimensions is shared between brains and models and accounts for a large fraction of brain-model correspondence in visual tasks \cite{chen2024}.}

\textcolor{replyblue}{Layer-wise profiles revealed a parallel with the posterior-to-anterior ordering expected from the visual hierarchy (Fig.~\ref{fig:hierarchy}) \cite{DiCarlo2012, Felleman1991,Kravitz2013}, but the mapping was not one-to-one. For example, while the ventral hub showed correspondence across multiple depths, the MT+/LOTC hub showed selective alignment restricted to specific stimulus categories. This is consistent with cortical representations drawing on features distributed across multiple stages and being modulated by stimulus content, rather than following a strictly feedforward depth-to-stage mapping of models.}

\textcolor{replyblue}{Language models exhibited a distinct profile, reflecting the nature of their input. Because these models processed tokenized captions--abstracted from pixel data--they did not follow the cortical depth gradient observed in vision models. Instead, alignment was concentrated in LOTC and adjacent temporal regions, driven primarily by biological-agent content. Layer-wise alignment rose early and plateaued, scaling with tokenizer alignment (Supplementary Note~\ref{subsection:tokenizer}). This pattern suggests a correspondence driven by coarse semantic content--closer to a ``bag-of-concepts'' organization than to progressively composed syntax \cite{yuksekgonul2023when}--indicating convergence on LOTC dimensions that are independent of low-level visual features.}

\textcolor{replyblue}{More generally, these findings align with observations that modern models tend to converge on similar latent representations across objectives and architectures as they scale \cite{chen2024, huh2024,jha2025}. One interpretation is that shared stimulus statistics, combined with efficiency constraints \cite{Friston2005}, favor the emergence of common representational axes that organize natural inputs by capturing structure essential for prediction and compression.}

\subsection*{Reliability and interpreting RSA magnitudes}

\textcolor{replyblue}{Across analyses, vision-model alignment closely tracked inter-subject alignment (Fig.~\ref{fig:spatial}e). RSA variants that differ in their noise sensitivity (e.g., within-subject RSA or alternative repetition-matching schemes) were also related by approximate power-law scaling (Supplementary Note~\ref{section:noise-ceiling}). This pattern is consistent with these measures indexing a common stimulus-locked representational geometry, differing primarily in how strongly measurement noise attenuates the observed correlations.}

\textcolor{replyblue}{This explains cases where model-brain RSA numerically exceeded IS-RSA in parcels with low-to-moderate reliability (e.g., posterior cingulate cortex; Supplementary Fig.~\ref{fig:extended_spatial_distribution}). IS-RSA correlates two noisy brain estimates, whereas model-brain RSA pairs one noisy estimate with a effectively noiseless model RDM. When reliability is limited, the noise from both sources in IS-RSA compounds the attenuation, lowering the observed correlation more than in model-brain RSA, even if both track the same underlying geometry.}

\textcolor{replyblue}{These results indicate that raw IS-RSA values may not always serve as an appropriate numeric ceiling for model-brain correspondence \cite{Nili2014,Diedrichsen2017,Walther2016}. Instead, RSA magnitudes are most informative when interpreted relative to reliability. Crucially, while adequate signal quality is a prerequisite for observing shared geometry--meaning the spatial extent of these maps inherently reflects the distribution of reliable fMRI signal--reliability alone cannot explain the topological organization we observed. The recovery of dissociable streams with distinct feature selectivities (e.g., ventral vs. lateral) confirms that this framework identifies specific information-processing networks, rather than merely tracking cortex-wide variations in signal quality. The empirical scaling relations provided here offer a practical calibration for separating these effects.}

\subsection*{Conclusions and scope}

\textcolor{replyblue}{In this work, we characterized a cortical network engaged during scene viewing at the level of representational geometry. By integrating complementary analyses of alignment, representational connectivity, and dimensionality decomposition, we linked three questions often treated separately: where shared stimulus information is expressed, how that information is routed across regions, and which stimulus properties organize representations within different network hubs. Together, these analyses provide a comprehensive perspective on how scene information is structured across the human cortex.}

\textcolor{replyblue}{This representational approach offers a general and adaptable methodology for identifying task-engaged networks and comparing information transformation across diverse systems. However, this flexibility entails a tradeoff: inferences concern representational organization rather than the precise mechanisms that implement it \cite{Baker2022,Cao2022}. Because these comparisons rely on global similarity structure, computational dimensions that are subtle but behaviorally relevant may contribute weakly to the measured geometry, or be missed entirely if not captured by the chosen metric or feature sampling \cite{lampinen2024}. Consequently, these results should be interpreted as context-dependent--varying with stimulus sampling, task demands, and measurement reliability \cite{Dujmovic2024inferring}--motivating future work to causally isolate the specific stimulus factors driving these alignments.}

\textcolor{replyblue}{Despite these limitations, this framework provides a practical route for leveraging large-scale data to move beyond simply asking whether two systems encode information similarly. By decomposing shared geometry to isolate driving features, the approach helps identify candidate factors and interactions that explain \textit{what} information is shared across cortex and \textit{how} it is distributed. As dataset scale and reliability increase, these tools can bridge the gap between network-level descriptions and experiments probing how distributed cortical systems support perception and flexible behavior in dynamic environments.}

\section{Methods}\label{methods}

\subsection{Functional-MRI data}

For the main analyses, we used the Natural Scenes Dataset (NSD) \cite{Allen2022}, a high-resolution 7T fMRI dataset with 8 participants performing an image recognition-memory task. On each trial, participants viewed a natural scene for 3s and reported whether it was old or new. Each participant completed 32–40 scanning sessions (750 trials per session), yielding up to $\sim30,000$ trials per participant with repeated presentations of images across sessions. Full acquisition and experimental details are reported in Allen et al.\ \cite{Allen2022}.

Analyses were based on the 1.0-mm volumetric preprocessing of the NSD dataset and the corresponding single-trial $\beta$-estimates (version 3) \cite{Allen2022}. We used GLM-denoised single-trial $\beta$-estimates \cite{GLMSingle2022}, as their image-related information content has been validated in multiple decoding and image reconstruction studies \cite{Takagi2023, Ozcelik2023, Scotti2023}. Voxels were grouped into cortical parcels using the HCP-MMP1.0 atlas \cite{Glasser2016}, which defines 180 cortical regions per hemisphere. We employed the pre-aligned parcel masks provided by the NSD authors to extract voxel activity within each cortical region \cite{Allen2022}. Analyses were performed using both the symmetric version of the atlas (combining hemispheres) and the asymmetric version (preserving hemisphere separation); a supplementary analysis confirms that symmetric parcels display highly correlated representational profiles (Supplementary Note~\ref{subsection:asymmetry}). In all figures, parcels are colored by macro-anatomical groups as defined in Huang et al. (2022) \cite{Huang2022}, to facilitate anatomical interpretation.

For each participant $p~\in~\mathcal{P}$, parcel $r~\in~\mathcal{R}$, and session $s~\in~\mathcal{T}p$, we define a response matrix $X_{p, r, s}~\in~\mathbb{R}^{n \times v_{p,r}}$ containing the voxel responses to each stimulus in that session. Here, $n$ denotes the number of trials (750 per session), and $v_{p,r}$ the number of voxels in parcel $r$ for participant $p$.

Comparative analyses were performed using fMRI data from the BOLD5000 \cite{Chang2019}, and THINGS \cite{Hebart2023} datasets. \textcolor{replyblue}{We selected these datasets because IS-RSA and KMCCA require a large number of matched stimuli across participants, which is often limited in larger-cohort naturalistic datasets.} Preprocessing and analysis followed the same procedure described above, with minor adaptations to accommodate differences in stimulus sampling and repetition structure. Details are provided in Supplementary Note~\ref{section:comparative}.

All data analysed are fully de-identified and publicly available. The original NSD collection was approved by the University of Minnesota Institutional Review Board (Allen et al.\cite{Allen2022}). The THINGS dataset was collected under NIH Institutional Review Board approval (Hebart et al.\cite{Hebart2023}), and BOLD5000 under the Institutional Review Board of Carnegie Mellon University (Chang et al.\cite{Chang2019}). No new human-subjects data were acquired for this work.

\subsection{Model Features Extraction}

To compare how deep learning models encode visual and linguistic stimuli relative to fMRI responses, we selected vision and language models previously used by Huh et al.\ (2024) in studies of representational convergence across modalities \cite{huh2024}. Model selection was conducted prior to analysis, balancing computational constraints and the goal of covering diverse training objectives and model scales. A full list of models is provided in Table~\ref{tab:foundation_models}.

We selected 17 vision models spanning families with distinct training regimes, including ViT AugReg\cite{steiner2022how}, CLIP\cite{Radford2021, Cherti2023}, DINOv2\cite{oquab2024dinov}, and MAE\cite{mae}. All architectures were based on sequential Vision Transformer (ViT) blocks\cite{dosovitskiy2021an}. Each NSD image was processed through the backbone of each model, and activations were extracted from the output of each ViT block (see Fig.~\ref{fig:methodology}A).

For language models, we selected 14 transformer-based encoders from the BLOOMZ\cite{muennighoff-etal-2023-crosslingual}, Gemma 2\cite{gemmateam2024gemma2improvingopen}, LLaMA\cite{touvron2023llama, openlm2023openllama}, and LLaMA 3\cite{grattafiori2024llama} families\cite{Vaswani2017}. For each NSD image, we used text captions annotated by trained raters and originally sourced from the MS-COCO dataset\cite{mscoco}. Captions were processed through each model’s encoder, and activations were extracted at the output of each transformer block.

Feature extraction was performed using the Transformers\cite{wolf2020transformers} and TIMM\cite{rw2019timm} libraries, adapting procedures from Huh et al.\cite{huh2024}. All models were implemented in PyTorch\cite{Pytorch2019} and retrieved from the {Hugging Face Model Hub}.

For each model $m \in \mathcal{M}$, layer $l \in \mathcal{L}_m$, and stimulus set $s \in \mathcal{T}$ (images or their corresponding captions), we constructed a matrix of layer activations $Z_{m, l, s} \in \mathbb{R}^{n_s \times v_m}$, where $n_s$ is the number of stimuli in set $s$ and $v_m$ is the model’s embedding size, which is constant across all layers in the selected models. For each model and stimulus set, the activations across layers define a sequence of latent representations, reflecting the hierarchical transformation of stimulus content within each model.

\begin{table}[ht]
\captionsetup{skip=6pt}
\centering
\small
\renewcommand{\arraystretch}{1.15}
\begin{tabular}{c c c l c c c} 
\toprule
Modality & Family & Training Regime & \multicolumn{1}{c}{Model} & \# Params & \# Blocks & Embedding size \\ 
\midrule
\multirow{17}{*}{Vision} 
  & \multirow{4}{*}{\begin{tabular}[c]{@{}c@{}}ViT\\(AugReg)\cite{steiner2022how}\end{tabular}}  
  & \multirow{4}{*}{\begin{tabular}[c]{@{}c@{}}Supervised\\(21 K classes)\end{tabular}}
      & ViT (AugReg) Tiny  & 10M  & 12 & 192  \\ 
  & & & ViT (AugReg) Small & 30M  & 12 & 384  \\ 
  & & & ViT (AugReg) Base  & 103M & 12 & 768 \\ 
  & & & ViT (AugReg) Large & 326M & 24 & 1024 \\ 
  \cline{2-7}
  & \multirow{6}{*}{\begin{tabular}[c]{@{}c@{}}CLIP\\(Vision)\cite{Radford2021, Cherti2023}\end{tabular}} 
  & \multirow{3}{*}{\begin{tabular}[c]{@{}c@{}}Contrastive\\Image–Text\end{tabular}}
      & CLIP (ViT) Base   & 86M  & 12 & 768  \\ 
  & & & CLIP (ViT) Large  & 304M & 24 & 1024 \\ 
  & & & CLIP (ViT) Huge   & 632M & 32 & 1280 \\ 
  \cline{3-7}
  & & \multirow{3}{*}{\begin{tabular}[c]{@{}c@{}}Contrastive I-T +\\fine‑tuning (12 K)\end{tabular}}
      & CLIP (ViT) Base ft‑12k  & 95M  & 12 & 768  \\ 
  & & & CLIP (ViT) Large ft‑12k & 315M & 24 & 1024 \\ 
  & & & CLIP (ViT) Huge ft‑12k  & 646 M & 32 & 1280 \\ 
  \cline{2-7}
  & \multirow{4}{*}{DINOv2\cite{oquab2024dinov}} & \multirow{4}{*}{Self‑Supervised}        
      & DinoV2 (ViT) Small & 22M  & 12 & 384  \\ 
  & & & DinoV2 (ViT) Base  & 87M  & 12 & 768  \\ 
  & & & DinoV2 (ViT) Large & 304M & 24 & 1024 \\ 
  & & & DinoV2 (ViT) Giant & 1.1B & 40 & 1536 \\ 
  \cline{2-7}
  & \multirow{3}{*}{MAE\cite{mae}} & \multirow{3}{*}{\begin{tabular}[c]{@{}c@{}}Self‑Supervised\\(Masked AE)\end{tabular}}          
      & MAE (ViT) Base   & 86M  & 12 & 768  \\ 
  & & & MAE (ViT) Large  & 303M & 24 & 1024 \\ 
  & & & MAE (ViT) Huge   & 631M & 32 & 1280 \\ 
\midrule
\multirow{14}{*}{Language} 
  & \multirow{5}{*}{BLOOMZ\cite{muennighoff-etal-2023-crosslingual}}                    & \multirow{5}{*}{\begin{tabular}[c]{@{}c@{}}Causal LM +\\Instruction FT\end{tabular}} 
      & BloomZ (560M) & 559M & 25 & 1024 \\ 
  & & & BloomZ (1b1)  & 1.1B & 25 & 1536 \\ 
  & & & BloomZ (1b7)  & 1.7B & 25 & 2048 \\ 
  & & & BloomZ (3B)   & 3B   & 31 & 2560 \\ 
  & & & BloomZ (7b1)  & 7.1B & 31 & 4096 \\ 
  \cline{2-7}
  & \multirow{2}{*}{Gemma 2\cite{gemmateam2024gemma2improvingopen}} & \multirow{2}{*}{Causal LM}
      & Gemma 2 (2B) & 2.6B & 27 & 2304 \\ 
  & & & Gemma 2 (9B) & 9.2B & 43 & 3584 \\ 
  \cline{2-7}
  & \multirow{5}{*}{LLaMA\cite{touvron2023llama, openlm2023openllama}} & \multirow{5}{*}{Causal LM}                
      & OpenLlama (3B)  & 3.4B & 27 & 3200 \\ 
  & & & OpenLlama (7B)  & 6.7B & 33 & 4096 \\ 
  & & & OpenLlama (13B) & 13B & 41 & 5120 \\ 
  \cline{4-7}
  & & & HuggyLLaMa (7B) & 6.7B & 33 & 4096 \\ 
  & & & HuggyLLaMa (13B)& 13B & 41 & 5120 \\ 
  \cline{2-7}
  & \multirow{2}{*}{LLaMA 3\cite{grattafiori2024llama}} & \multirow{2}{*}{Causal LM}
      & Llama 3 (8B)   & 8B  & 33 & 4096 \\ 
  & & & Llama 3.1 (8B) & 8B  & 33 & 4096 \\ 
\bottomrule
\end{tabular}
\caption{Vision and language models used in the analyses.}
\label{tab:foundation_models}
\end{table}

\subsection{Alignment Measure}

We used Representational Similarity Analysis (RSA)\cite{Kriegeskorte2008} to quantify the correspondence between brain parcels and model layers geometries. For each system, the representational dissimilarity matrix (RDM) $D(X)$ was defined as $D(X)_{ij} = 1 - \rho(x_i, x_j)$, where $x_i$ and $x_j$ are rows of $X$ and $\rho$ is the Pearson correlation. The RSA score between two systems $X$ and $Y$ was the correlation between the upper triangles of their RDMs:
\begin{equation}
    \operatorname{RSA}(X, Y) = \rho\left( \operatorname{vec}_u D(X),\; \operatorname{vec}_u D(Y) \right)
\end{equation}.

Main analyses were replicated with Spearman correlation and Centered Kernel Alignment (CKA)\cite{kornblith2019cka}, yielding consistent results (see Supplementary Note~\ref{sec:metric-comparison}). Details of the matrix reformulation that enabled efficient, large-scale RSA computations are provided in Supplementary Note~\ref{section:implementation}.

\subsection{Inter-Subject alignment}

Inter-subject RSA was computed using the subset of 1,000 images viewed by all NSD participants (the \textit{shared1000} set)\cite{Allen2022}. Each image in this subset was presented up to three times per participant, with corresponding trial indices across sessions. Trials were uniquely identified by participant, image ID, and repetition index $k \in {1,2,3}$. For each pair of participants $p, q \in \mathcal{P}$ and cortical parcels $r, r' \in \mathcal{R}$, we extracted the corresponding voxel responses $X_{p,r}$ and $X_{q,r'}$ and computed RSA between matched trials.

Images shared across participants were always presented at the same trial positions for all participants, interleaved with unique images specific to each individual. To attenuate alignment driven by task or session structure rather than by stimulus processing, we applied a cyclic repetition shift when matching trials across participants. Specifically, for each image $i$ and repetition index $k$, the trial $(i, k)$ in participant $p$ was matched with the trial $(i, (k \bmod 3) + 1)$ in participant $q$, thereby preserving stimulus identity while disrupting repetition-locked confounds. This matching strategy and its impact are illustrated and compared in Supplementary Note~\ref{subsection:repetition-matching}. Let $\boldsymbol{i}_{pq}$ and $\boldsymbol{j}_{pq}$ denote the matched stimulus indices for participants $p$ and $q$, respectively. The inter-subject alignment between parcels $r$ and $r'$ was computed as:

\begin{equation}
\mathcal{A}^{\mathrm{IS}}_{p, q, r,r'} =
\operatorname{RSA}\left(
X_{p,r}[\boldsymbol{i}_{pq}, :],\;
X_{q,r'}[\boldsymbol{j}_{pq}, :]
\right).
\end{equation}

To compute a subject-level alignment measure for participant $p$, we aggregated alignment values with respect to all other participants:
\begin{equation}
\mathcal{A}^{\mathrm{IS}}_{p, r, r'} = 
\frac{1}{|\mathcal{P}|-1} \sum_{q \in \mathcal{P} \setminus \{p\}} 
\mathcal{A}^{\mathrm{IS}}_{p,q,r,r'}.
\end{equation}

Group-level alignment was then obtained by averaging subject-level measures across participants. Cortical surface maps (e.g., Fig.\hyperref[fig:spatial]{\ref*{fig:spatial}b}) display group-level alignment, specifically comparing the same parcels across participant pairs ($r = r'$). To benchmark the degree of shared geometry relative to individual consistency, we also compared inter-subject and within-subject alignment (see Supplementary Fig. \ref{fig:within-subject}), which yielded consistent results.

A parcel-by-parcel matrix of group-level inter-subject alignment values ($n_{\mathrm{rois}} \times n_{\mathrm{rois}}$) defined the whole-cortex connectivity network. We visualized the main network backbone and multiple major representation routes using an iterative minimum spanning tree (k-MST) approach on the group-averaged connectivity matrix. At each iteration, a standard MST was computed (with edge weights defined as $1-\mathrm{similarity}$), removing previously selected edges to ensure the inclusion of new pathways. The final backbone comprised the union of edges from all $k$ MSTs (see Supplementary Fig.~\ref{fig:whole_connectivity}), preserving multiple principal paths, avoiding arbitrary edge thresholding, and ensuring that all parcels remained connected while allowing for richer topology. MST computations were performed using the NetworkX and SciPy libraries\cite{Virtanem2020, hagberg2008exploring}.

Statistical significance was evaluated by comparing the observed alignment values to a null distribution generated by randomly shuffling stimulus labels (10,000 permutations, using a shared permutation scheme across all comparisons) with a two-tailed test. All p-values were FDR-corrected \cite{Benjamini1995}.

\subsection{Brain–Model alignment}

Brain–model representational alignment was computed using all available trials from NSD participants (i.e., not restricted to the set of shared images), with analyses performed separately by session to control for session-specific variance. For each participant $p \in \mathcal{P}$, model $m \in \mathcal{M}$, and session $s \in \mathcal{T}_p$, voxel responses from each cortical parcel $r \in \mathcal{R}$ were compared to model activations at each layer $l \in \mathcal{L}_m$ using RSA:

\begin{equation}
\mathcal{C}_{p, r, s, m}(l) = 
\operatorname{RSA}\bigl(
  X_{p, r, s},\;
  Z_{m, l, s}
\bigr).
\end{equation}

To compute a subject-level alignment measure for each model, we identified the layer with the largest absolute alignment for each session, preserving the sign of the value (denoted as $\operatorname{sign\,max}$ in Eq.~\ref{eqn:mb-align}). These peak values were then averaged across sessions:
\begin{equation}
\label{eqn:mb-align}
\mathcal{A}^{\mathrm{BM}}_{p, r, m} = 
\frac{1}{|\mathcal{T}_p|} \sum_{s \in \mathcal{T}_p} 
\operatorname*{sign\,max}_{l \in \mathcal{L}_m}\,\mathcal{C}_{p, r, s, m}(l).
\end{equation}
This procedure, standard in recent literature \cite{Schrimpf2020, huh2024}, ensures that alignment is captured independently of model depth. In cases-such as for language models-where correlations may be systematically negative, this approach preserves the interpretability of dissimilar representational geometries: negative alignment values indicate that stimuli which are close in one space (e.g., visual cortex) are systematically distant in another (e.g., language model embedding)\cite{Schutt2023}.

To identify where alignment peaked within the model hierarchy (e.g., Fig.~\ref{fig:hierarchy}b–c), we computed the normalized depth of the maximally aligned layer for each session and averaged across sessions:
\begin{equation}
d^{*}_{p, r, m} =
\frac{1}{|\mathcal{T}_p|} \sum_{s \in \mathcal{T}_p}
\frac{1}{|\mathcal{L}_m| - 1}
\arg\max_{l \in \mathcal{L}_m}
\mathcal{C}_{p, r, s, m}(l),
\end{equation}
where layers are indexed from 0 (first layer) to $|\mathcal{L}_m| - 1$ (last layer), so $d^{*}_{p, r, m} \in [0, 1]$. This procedure is standard for mapping the correspondence between brain regions and network hierarchy \cite{Caucheteux2022, Schrimpf2020}.
To obtain a subject-level measure for a set of models (e.g., all vision or language models), both the peak alignment and normalized depth measures were averaged across models. Group-level measures were then obtained by averaging subject-level values across participants.

To study how alignment changes across the model hierarchy, we constructed continuous alignment curves for each subject parcel and model by averaging layer-wise RSA across sessions. To compare models with different numbers of layers, linear interpolation was applied, normalizing layer indices to a depth variable $d \in [0, 1]$ ($d = 0$ for the first layer, $d = 1$ for the last layer):
\begin{equation}
\tilde{\mathcal{C}}_{p, r, m}(d) = 
\operatorname*{interp}_{l \to d}\left(
\frac{1}{|\mathcal{T}_p|} \sum_{s \in \mathcal{T}_p}
\mathcal{C}_{p, r, s, m}(l)
\right), \quad d \in [0,1].
\end{equation}

Alignment curves for each subject’s parcel across a set of models $\mathcal{M}$ (e.g., all vision models or all language models) were then obtained by averaging interpolated curves across models:
\begin{equation}
\tilde{\mathcal{C}}_{p, r, \mathcal{M}}(d) =
\frac{1}{|\mathcal{M}|} \sum_{m \in \mathcal{M}}
\tilde{\mathcal{C}}_{p, r, m}(d).
\end{equation}

Statistical significance was assessed using a permutation test with 10,000 stimulus-label shuffles. The null distribution was generated by applying the same set of permutations across all model layers and parcels, allowing direct comparison and group averaging. Two-tailed p-values were calculated by comparing the observed group-mean alignment to this null, and corrected for multiple comparisons using the Benjamini–Hochberg FDR procedure.

\subsection{Shared Component Extraction}

To identify the dominant dimensions underlying shared representational geometry across participants, we applied kernel multi-view canonical correlation analysis (KMCCA)~\cite{Hardoon2004MKCCA}. KMCCA was performed on the set of image trials shared by all participants in NSD, extracting a low-dimensional subspace that maximized correlation across participants’ RDMs for each hub.

We used the KMCCA implementation from \emph{mvlearn}\cite{perry2021mvlearn}. For each hub, we extracted the first two components for exploratory purposes, and assessed the influence of the first (dominant) component in control analyses. While our focus was on the dominant axis, further work is needed to interpret all statistically significant components. Image-level semantic annotations and category labels were derived from MS-COCO~\cite{mscoco}. For automated labeling of the scene-to-object gradient, we used Pixtral-12B~\cite{agrawal2024pixtral12b} (see Fig.~\hyperref[fig:components]{\ref*{fig:components}a}).

Further details of the KMCCA procedure, semantic projections, and partial RSA controls are provided in Supplementary Note~\ref{section:extended_components}.

\subsection{Statistics and Reproducibility}

\textcolor{replyblue}{Statistical testing and analysis were carried out in Python 3.10 using standard libraries (SciPy, Statsmodels). No statistical methods were used to pre-determine sample size; sample sizes were determined by availability in the public datasets: Natural Scenes Dataset ($N=8$), BOLD5000 ($N=4$), and THINGS-fMRI ($N=3$). These sizes are standard for deep-sampling fMRI studies where individual-subject reliability is high. All subjects and sessions from these datasets were included in the analysis. To assess the statistical significance of RSA maps (e.g., Fig.~\ref{fig:spatial}a-c), a two-tailed permutation test ($n=10,000$) was used to determine if RSA values significantly differed from a null distribution. For all other NSD group-level results (Figs. 1--5), significance was assessed using two-tailed paired t-tests ($N=8$) unless otherwise specified. In all cases, $p$-values were corrected for multiple comparisons using the False Discovery Rate (FDR) or Bonferroni method, with significance defined as $p < 0.05$.}

\section*{Data and code availability}
Functional-MRI data and image stimuli were obtained from the Natural Scenes Dataset (accessed via \href{https://registry.opendata.aws/nsd}{https://registry.opendata.aws/nsd}), BOLD5000 (\href{https://doi.org/doi:10.18112/openneuro.ds001499.v1.3.0}{10.18112/openneuro.ds001499.v1.3.0} and \href{https://doi.org/10.1184/R1/c.5325683}{10.1184/R1/c.5325683}), and THINGS-fMRI (\href{https://doi.org/10.25452/figshare.plus.c.6161151.v1}{10.25452/figshare.plus.c.6161151.v1} and \href{http://doi.org/10.17605/osf.io/jum2f}{10.17605/osf.io/jum2f}). 
Image metadata were extracted from MS-COCO (\href{https://cocodataset.org}{https://cocodataset.org}). Model checkpoints analyzed in this work are available on Hugging Face (\href{https://huggingface.co/collections/pablomm/convergent-transformations-6808f7de248fa9674acac588}{https://huggingface.co/collections/pablomm/convergent-transformations-6808f7de248fa9674acac588}). \textcolor{replyblue}{Precomputed derivative data generated from these raw sources are available in Figshare \href{https://doi.org/10.6084/m9.figshare.30753239}{10.6084/m9.figshare.30753239}.
All resources were publicly accessible in January 2026.}

Code to reproduce analyses is available at \href{https://github.com/memory-formation/convergent-transformations}{https://github.com/memory-formation/convergent-transformations}.\\

\section*{Acknowledgments}
We thank J.\ García-Arch, M. Domínguez-Orfila, D.\ Pacheco-Estefan, C.\ Baldassano, and E.\ Cámara for helpful comments and discussions.
This work was supported by the Spanish Ministerio de Ciencia, Innovación y Universidades, which is part of Agencia Estatal de Investigación (AEI), through the project PID2022\ -\ 140426NB (Co-funded by European Regional Development Fund. ERDF, a way to build Europe). We thank CERCA Programme/Generalitat de Catalunya for institutional support.

\section*{Author contributions}
P.M. and L.F. conceived the project, formulated the methodology and drafted the manuscript. P.M. performed the analyses. L.F. secured funding and supervised the work. Both authors revised and approved the final manuscript.

\section*{Competing interests}
\textcolor{replyblue}{The authors declare no competing interests.}

\bibliographystyle{unsrtnat}
\bibliography{references}

\clearpage

\include{supplementary}

\end{document}

%% file: supplementary.tex
\renewcommand{\thesection}{\arabic{section}}
\renewcommand{\thesubsection}{\arabic{subsection}} 
\renewcommand{\thefigure}{\arabic{figure}}     
\renewcommand{\thetable}{\arabic{table}} 
\renewcommand{\theequation}{\arabic{equation}}

\captionsetup[figure]{name=Supplementary Figure}

\setcounter{section}{0}
\setcounter{subsection}{0}
\setcounter{figure}{0}
\setcounter{table}{0}
\setcounter{equation}{0}

\titleformat{\subsection}
  {\normalfont\large\bfseries} 
  {}                           
  {0pt}                        
  {}                           

\begin{center}
    \Large\bfseries Supplementary Material
\end{center}

\section*{\centering{Extended figures}}

\begin{figure}[hbtp]
    \centering
    \includegraphics[width=\textwidth]{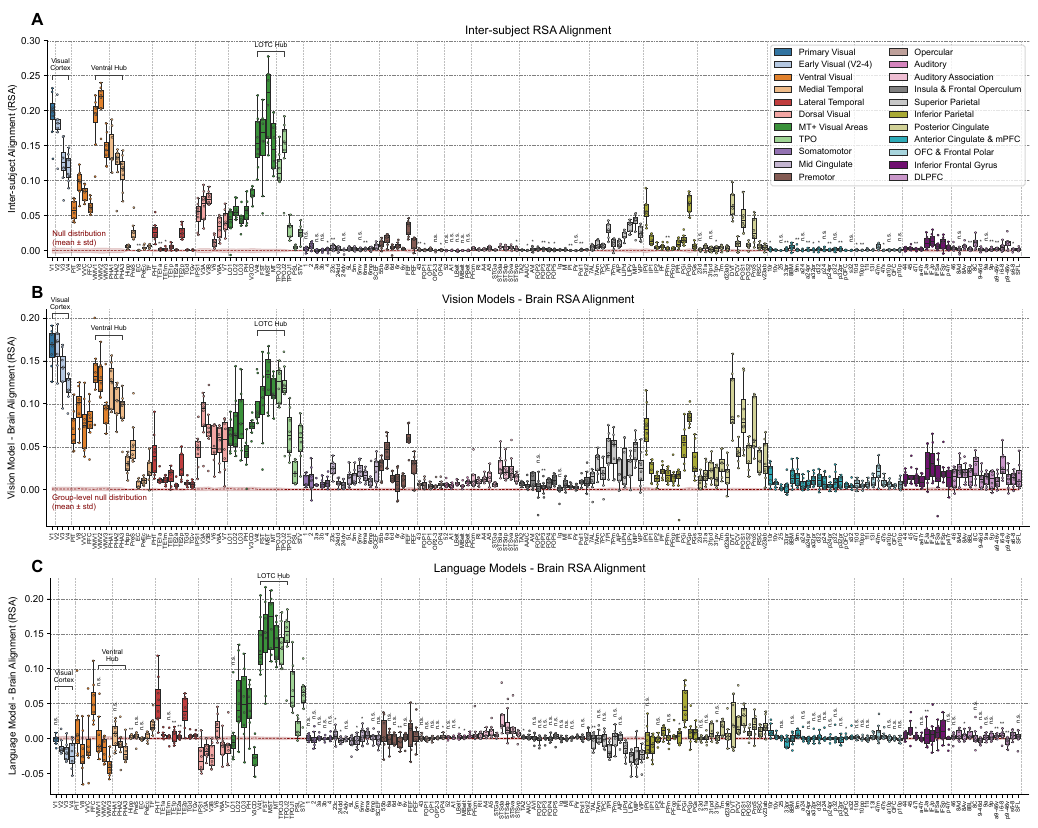}
    \caption{\textbf{Detailed parcel-level alignment for all modalities.}
Box plots show representational alignment scores computed for each of the 180 cortical parcels of the symmetric HCP atlas \cite{Glasser2016}, organized by macro-anatomical groups \cite{Huang2022}. Each box represents the distribution of alignment scores across the eight participants ($N=8$). The red line and shaded area in each panel denote the mean and standard deviation of the null distribution, respectively, estimated via permutation testing.
\textbf{(A)}~Inter-subject alignment (IS-RSA).
\textbf{(B)}~Brain-to-vision-model alignment.
\textbf{(C)}~Brain-to-language-model alignment.
This figure provides a detailed view of the results summarized in the main text, showing the parcel-by-parcel variability and confirming the concentration of high alignment within the Early Visual, Ventral, and LOTC hubs.
}
    \label{fig:extended_spatial_distribution}
\end{figure}

\begin{figure}[p]
    \centering
    \includegraphics[width=\textwidth]{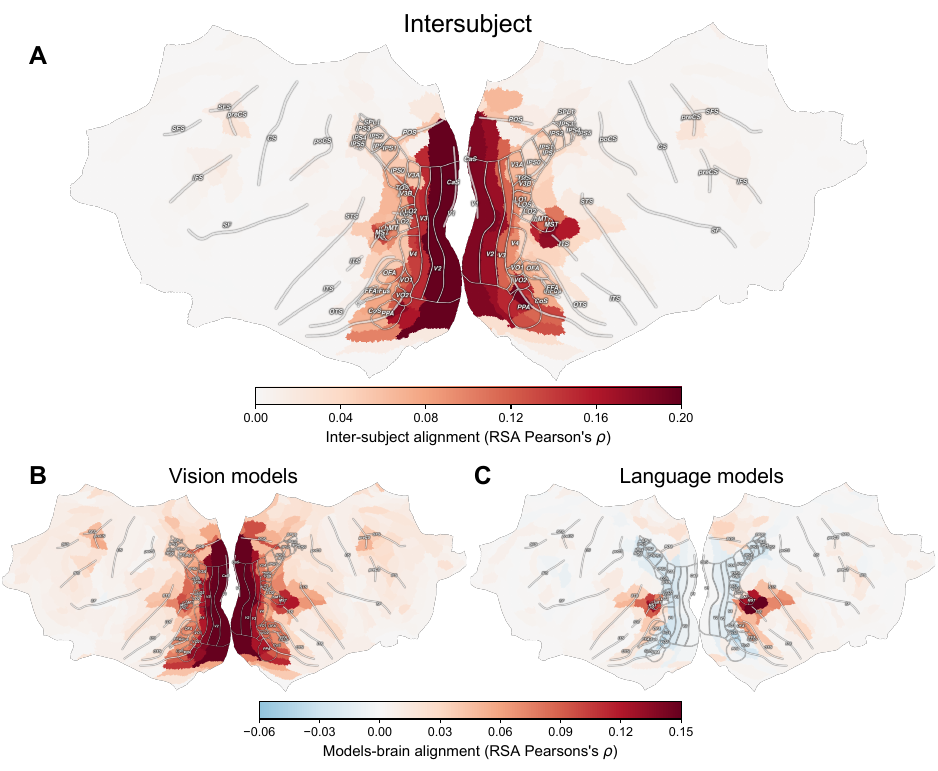}
    \caption{\textbf{Cortical flat map projections of representational alignment.}
To provide a comprehensive view of their spatial distribution, group-level alignment scores are projected onto a flattened cortical surface. The figure shows \textbf{(A)}~IS-RSA alignment, \textbf{(B)}~vision-model alignment, and \textbf{(C)}~language-model alignment. Values represent group-averaged RSA scores (Pearson's~$\rho$, $N=8$ from the NSD dataset), and major sulci are labeled for anatomical orientation. This visualization makes the full extent of the alignment patterns clear, particularly highlighting the widespread negative alignment (cool colors) between language models and the ventral visual stream, in contrast to the positive alignment seen for inter-subject and vision-model comparisons.
}
    \label{fig:cortical_surfaces}
\end{figure}

\begin{figure}[p]
    \centering
    \includegraphics[width=\linewidth]{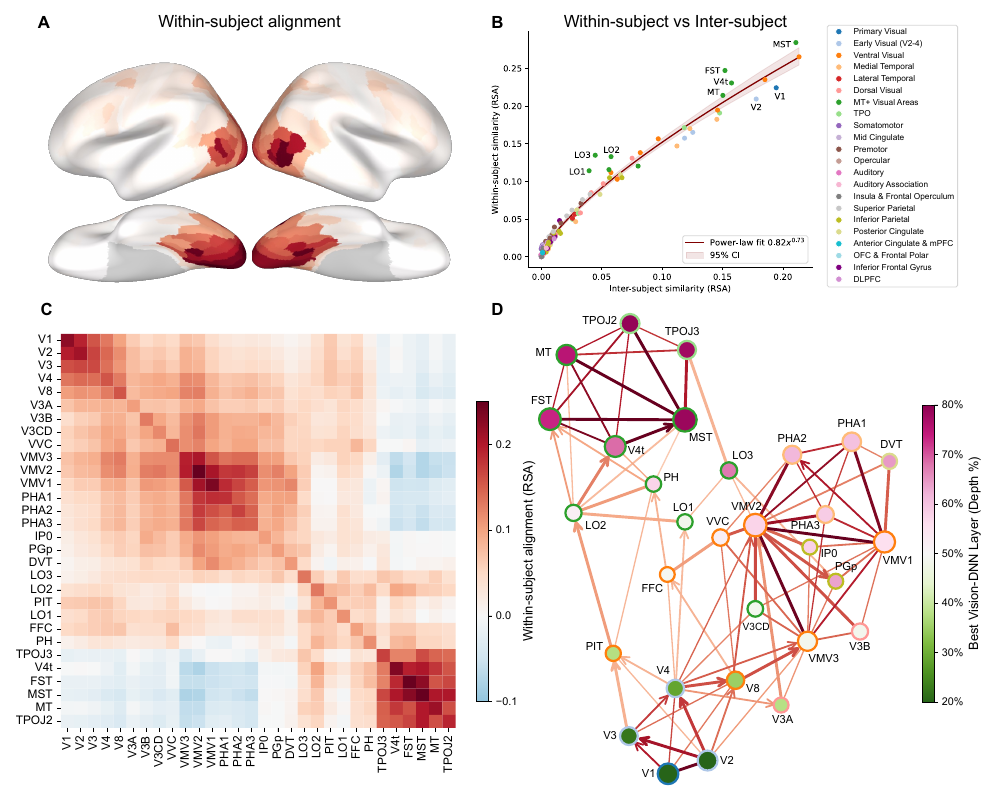}
     \caption{
     \textbf{Comparison of within-subject and inter-subject representational analyses.}
\textbf{(A) Cortical map of within-subject alignment} (RSA, Pearson's~$r$). This was computed analogously to the inter-subject analysis by comparing RDMs from different trial repetitions within each participant, and then averaging across participants ($N=8$). The map reveals a similar spatial distribution to the inter-subject version, with a posterior-to-anterior gradient and three distinct hubs.
\textbf{(B) Parcel-wise comparison of within-subject and inter-subject alignment}. The two measures are tightly correlated across the cortex, following a power-law relationship (fit shown in red; $R^2=0.98$). This indicates that brain regions that are consistent within an individual are also the ones that are consistent across individuals.
\textbf{(C) Within-subject representational connectivity matrix}. Computed by correlating RDMs between all pairs of parcels within each participant, this analysis reveals the same three-hub block structure discovered using the inter-subject measurements.
\textbf{(D) Directed connectivity graph of the within-subject data}, pruned using a 3-minimum spanning tree to highlight the network backbone. Edge color and width encode the inter-region alignment (RSA value), while node color represents each parcel's peak alignment depth with vision models. Directionality is inferred from this depth measure based on a low-to-high sequential hierarchy, with arrows pointing from shallower- to deeper-aligning regions. The graph confirms the same information flow as the inter-subject analysis, with two streams emerging from early visual cortex: a medial-ventral stream to the Ventral hub and a lateral-dorsal stream to the LOTC hub.}
    \label{fig:within-subject}
\end{figure}

\begin{figure}[p]
    \centering
    \includegraphics[width=0.78\textwidth]{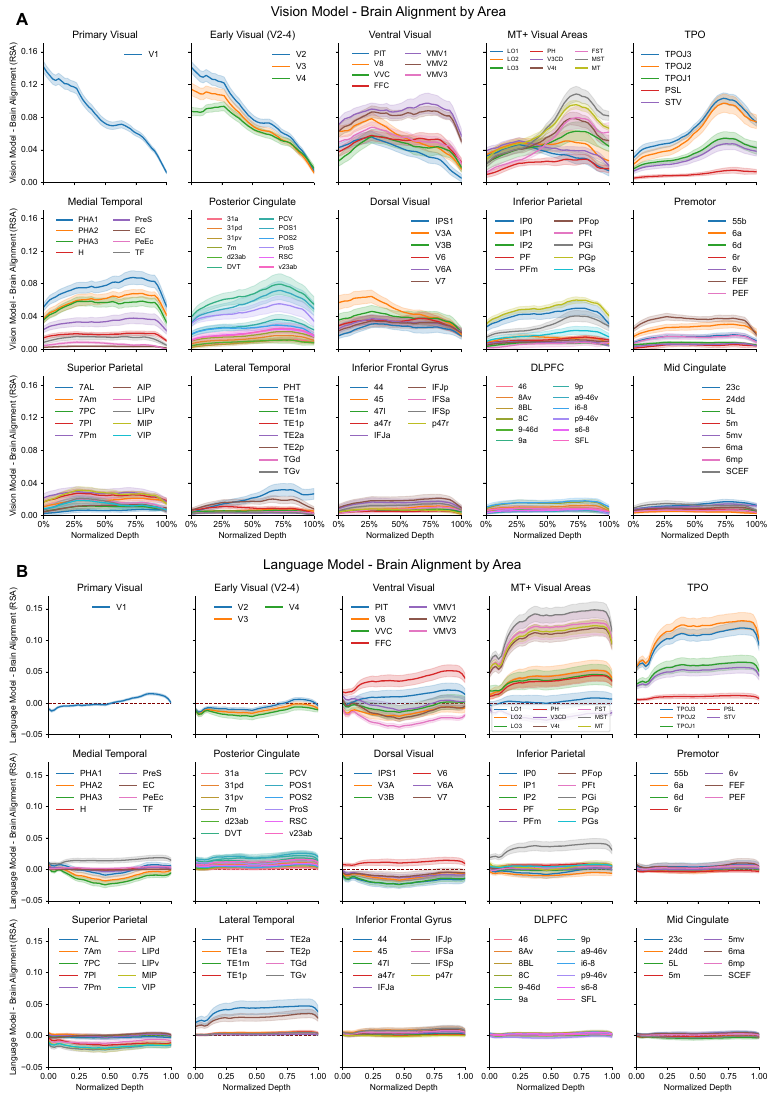}
    \caption{\textbf{Detailed layer-wise alignment profiles across cortical parcels.}
Detailed parcel-by-parcel view of the brain-model alignment curves summarized in the main text (Fig.~\ref{fig:hierarchy}), showing results for the 15 macro-anatomical groups with the highest inter-subject alignment. Only parcels that were themselves statistically significant in the inter-subject analysis are displayed. For each modality, the plotted alignment curves represent the RSA score averaged across all models of that type. Each individual line shows the mean alignment for a single parcel across participants ($N=8$), with shaded areas indicating the SEM.
\textbf{(A)~Brain-to-vision-model alignment}. This detailed view confirms that the distinct alignment profiles--decreasing, distributed, and increasing--are characteristic of the main hubs and often extend to anatomically adjacent parcels (e.g., the increasing profile of the LOTC hub is also observed in nearby STS regions). In contrast, areas with lower overall correspondence, such as in the prefrontal cortex, tend to show weak alignment across all model layers without a clear hierarchical preference.
\textbf{(B)~Brain-to-language-model alignment}. This view confirms that positive alignment is restricted to the LOTC hub and a small number of anatomically proximal parcels, all of which consistently exhibit the same step-like alignment profile.
}
    \label{fig:extneded_roi_curves}
\end{figure}

\begin{figure}[p]
    \centering
    \includegraphics[width=\textwidth]{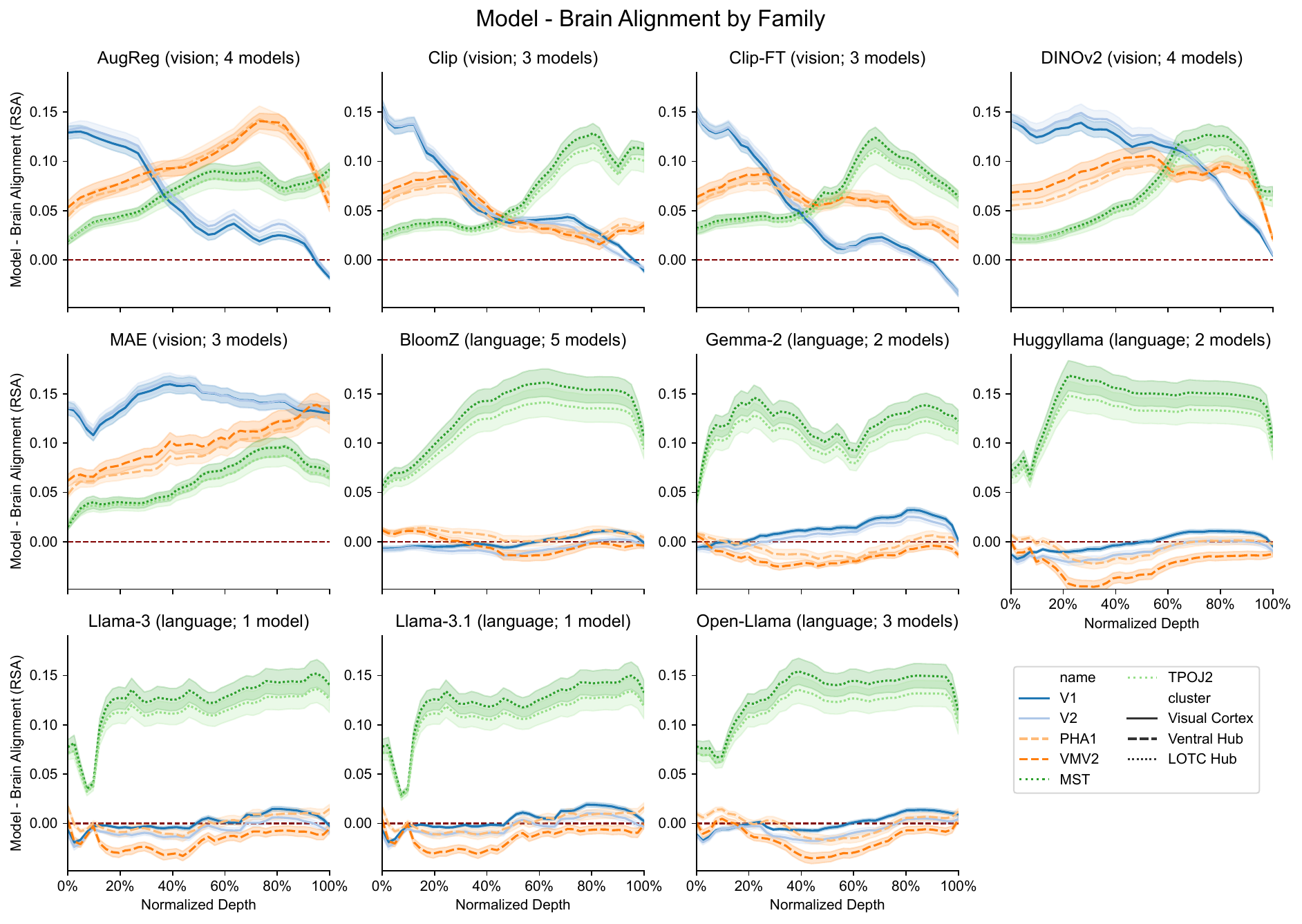}
\caption{\textbf{Brain-model alignment profiles for different model families.}
Layer-wise alignment (RSA, Pearson's~$r$) is shown for representative parcels from the three main hubs, computed separately for each family of models (see Table~\ref{tab:foundation_models} for a full list). Each curve represents the mean alignment across participants ($N=8$), with shaded areas indicating the SEM.
\textbf{Vision Models:} Most vision model families (AugReg, CLIP, DINOv2) reproduce the characteristic hierarchical profiles: a decreasing alignment with Early Visual Cortex, a distributed alignment with the Ventral Hub, and an increasing alignment with the LOTC Hub. The Masked Autoencoder (MAE) models are a notable exception; consistent with their reconstructive training objective, they maintain a high alignment with Early Visual Cortex across all layers.
\textbf{Language Models:} Despite differences in architecture, training data, and fine-tuning (e.g., instruction tuning in BLOOMZ), all language model families exhibit a similar pattern. Alignment is negligible or negative for Early Visual and Ventral hub parcels, while all families show the same characteristic step-function profile for parcels in the LOTC hub.}
    \label{fig:extneded_family_curves}
\end{figure}

\begin{figure}[p]
    \centering
    \includegraphics[width=\textwidth]{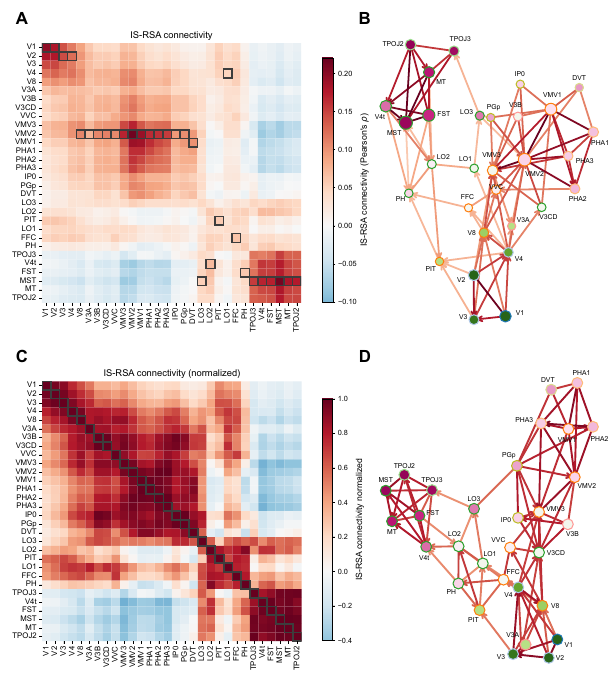}
    \caption{
Noise-ceiling normalization of IS-RSA representational connectivity.
(\textbf{A}) IS-RSA representational connectivity matrix (Pearson’s $\rho$) for the 30 parcels with highest inter-subject alignment. Black squares indicate, for each parcel (column), the parcel with which it shows maximal between-parcel IS-RSA connectivity.
(\textbf{B}) Backbone graph derived from (\textbf{A}) using three iterations of a minimum-spanning-tree procedure.
(\textbf{C}) Noise-ceiling–normalized connectivity matrix, obtained by dividing each edge by the geometric mean of the corresponding diagonal self-alignment estimates, setting the diagonal to 1.
(\textbf{D}) Backbone graph extracted from the normalized matrix in (\textbf{C}). The preserved topology indicates that the inferred dual-route architecture is robust to parcel-wise differences in measurement reliability.
}
    \label{fig:noise-ceil-connectivity}
\end{figure}

\begin{figure}[p]
    \centering
    \includegraphics[width=0.79\textwidth]{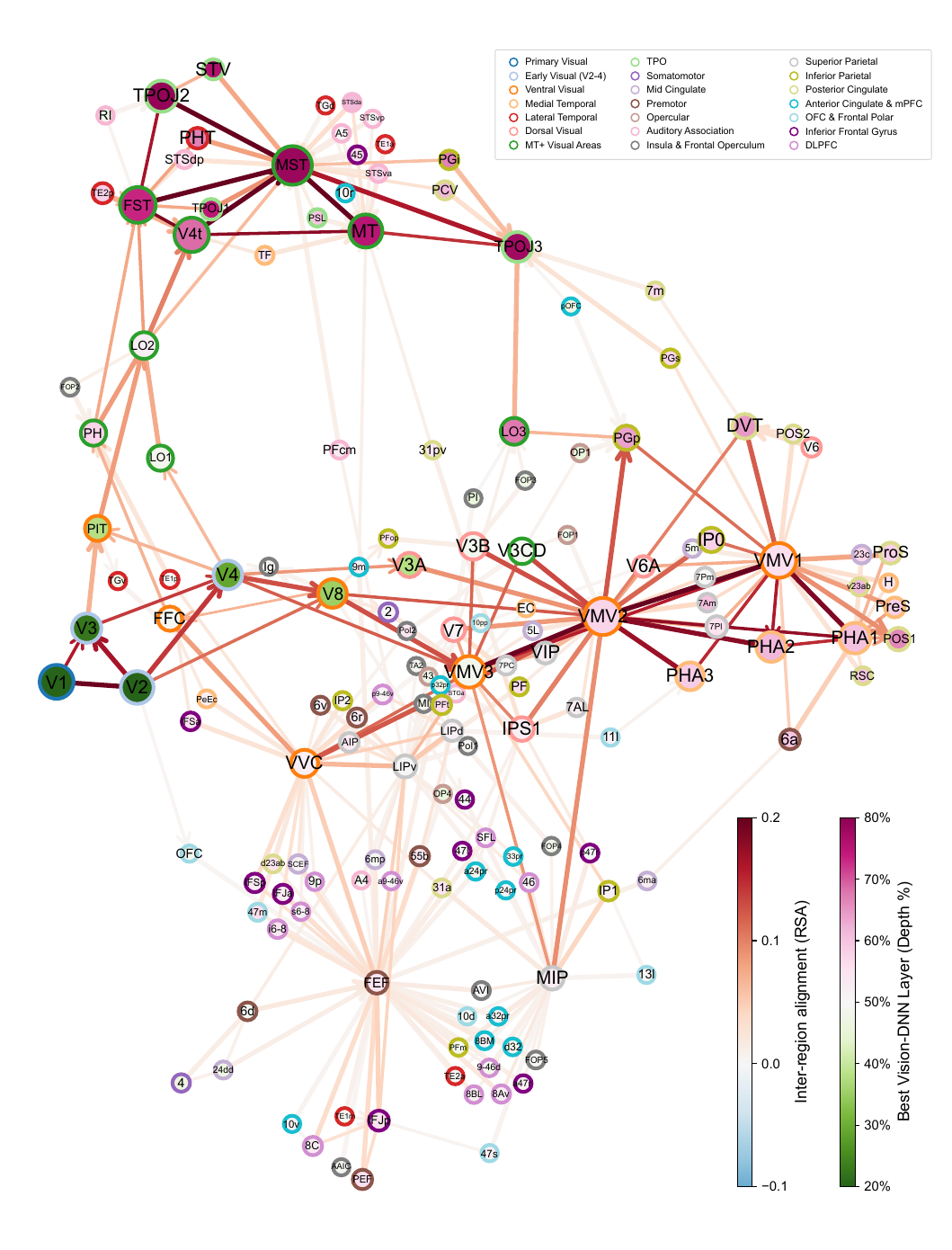}
    \caption{\textbf{Whole-cortex representational connectivity network.} Representational connectivity network among the 157 cortical parcels that exhibited statistically significant inter-subject alignment. The network is pruned using a 2-minimum spanning tree to highlight the strongest connections. To minimize confounds from anatomical variability, connectivity was computed on a within-subject basis ($N=8$); for each participant, we correlated the RDM of every parcel with every other parcel across different repetitions of the same stimuli, then averaged the resulting connectivity matrices. Edge color encodes the strength of this inter-region alignment (RSA), while edge width indicates the spanning tree order (1st or 2nd) to highlight the most central pathways. Node color represents each parcel's peak alignment depth with vision models, providing a proxy for its hierarchical position. Arrows are added to the pruned edges to denote a statistically significant difference in this depth between connected parcels (paired $t$-test, $t(7)$, FDR-corrected $p < 0.05$), indicating the putative direction of information flow.
The graph confirms the two primary visual processing streams identified in the main text. A medial-ventral stream connects early visual areas to core ventral stream regions involved in scene and object processing. In parallel, a lateral-dorsal stream connects early visual areas to the LOTC hub, including motion-sensitive areas (MT+ complex) and higher-order regions in the superior temporal sulcus (STS) and temporoparietal junction (TPOJ). The analysis also reveals key bridges between these streams, with nodes such as LO3 and PGp linking the ventral and LOTC hubs. Furthermore, despite weaker overall alignment, frontal and parietal regions like the Frontal Eye Fields (FEF) and the Medial Intraparietal Area (MIP) emerge as distinct hubs, suggesting their integration into this large-scale, stimulus-driven network.
}
    \label{fig:whole_connectivity}
\end{figure}

\begin{figure}[p]
    \centering
    \includegraphics[width=\textwidth]{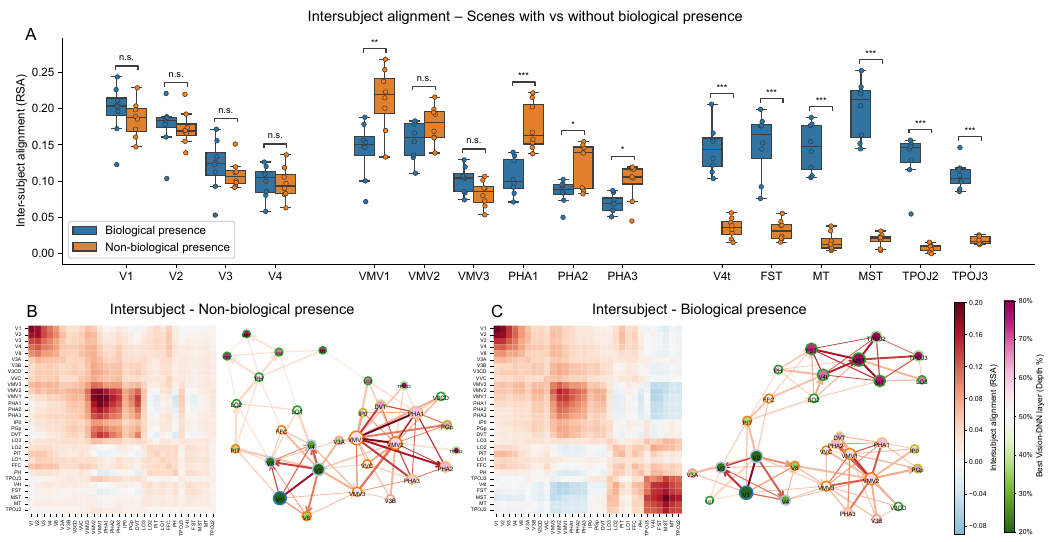}
     \caption{\textbf{Biological content selectively drives LOTC alignment and lateral‐stream connectivity.}
Scenes were split into those that contained biological agents (people or animals; 65.3\% of the shared image set) and those that did not, and all inter-subject analyses were recomputed on each subset.
\textbf{(A)} Parcel-wise inter-subject RSA within the three hubs.  
Blue: scenes with biological agents; orange: scenes without them.  Early Visual parcels (V1–V4) show no reliable difference, Ventral parcels (VMV1–PHA3) align more for non-biological scenes, whereas LOTC parcels (V4t–TPOJ3) align almost exclusively for biological scenes.  Stars mark two-tailed paired $t$-tests across eight participants ($^{*}p<0.05$, $^{**}p<0.01$, $^{***}p<0.001$, FDR-corrected).
\textbf{(B,C)} Group-level representational-connectivity matrices (left) and backbone graphs (right) for scenes \emph{without} (\textbf{B}) and \emph{with} (\textbf{C}) biological agents. 
Backbones were extracted with a three-iteration minimum-spanning-tree procedure.  
Node colour encodes each parcel’s peak vision-model layer (early $\rightarrow$ late: green $\rightarrow$ purple), node size scales with within-parcel IS-RSA, edge hue shows pairwise RSA, and edge width denotes the iteration (1–3) at which the edge entered the spanning tree.  
The lateral stream linking Early Visual Cortex to LOTC is largely absent for non-biological scenes (B) but re-emerges when biological agents are present (C), mirroring the parcel-level results in (A).}
    \label{fig:split_analysis}
\end{figure}

\begin{figure}[p]
    \centering
    \includegraphics[width=\textwidth]{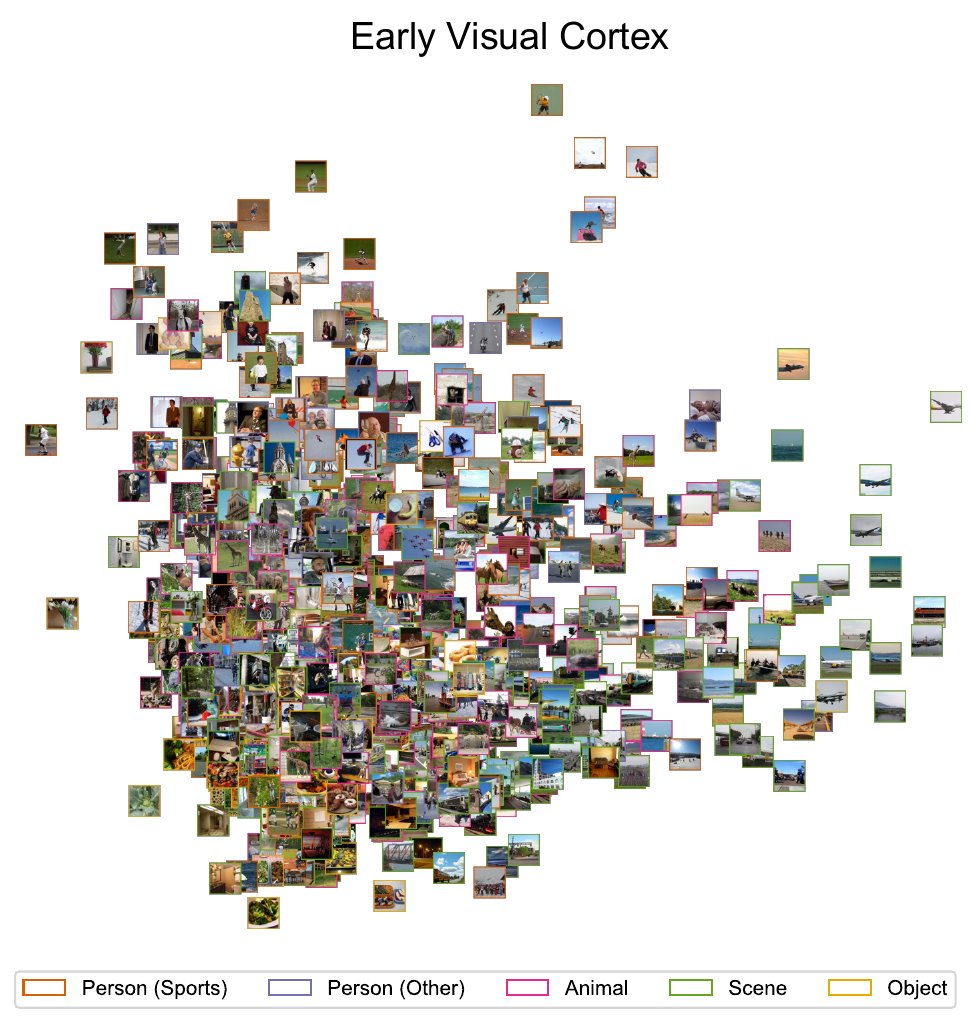}
\caption{\textbf{Image-based atlas of the dominant representational dimensions in Early Visual Cortex.}
The first two shared representational components for the Early Visual Cortex hub, derived using KMCCA on the responses of all participants to the shared image set. Each point is an individual stimulus image, and its colored border indicates its general semantic category (see legend). The layout reveals no clear clustering by semantic category. Instead, images that are close to each other in this shared space often have similar low-level visual properties (e.g., color palettes, textures, or global shapes), confirming that the geometry of this hub is primarily driven by visual similarity rather than abstract content.
}
    \label{fig:atlas_evc}
\end{figure}

\begin{figure}[p]
    \centering
    \includegraphics[width=0.9\textwidth]{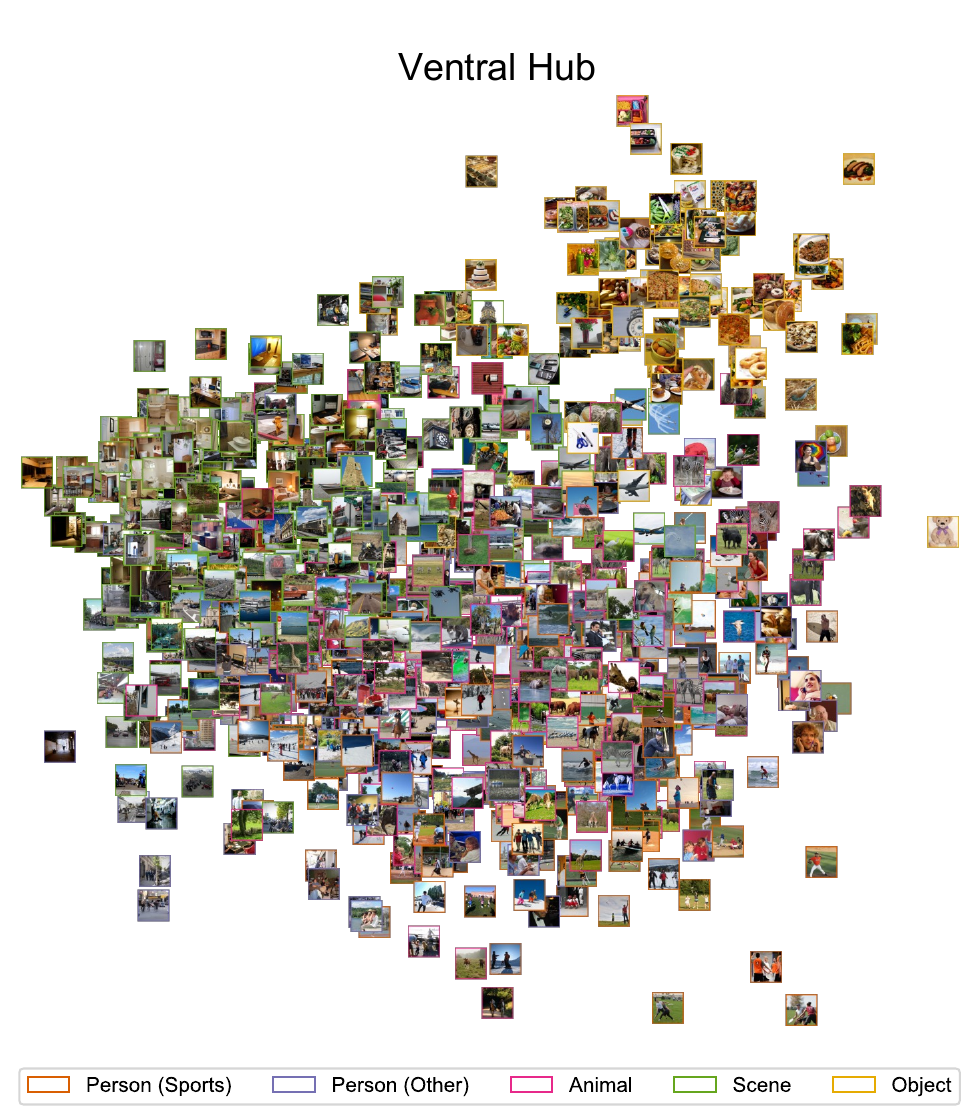}
\caption{\textbf{Image-based atlas of the dominant representational dimensions in the Ventral Hub.}
The first two shared representational components for the Ventral hub, derived using KMCCA. Each point is an individual stimulus image. The layout reveals an organizing principle: a smooth gradient from scene-dominated images to object-dominated images. For example, the upper-left quadrant contains images of indoor spaces, the lower portion contains outdoor scenes, and the upper-right quadrant contains images where a single object is the primary focus. This evidences that the dominant representational axis of this hub recapitulates the classic scene-to-object gradient of ventromedial cortex.
}
    \label{fig:atlas_ventral}
\end{figure}

\begin{figure}[p]
    \centering
    \includegraphics[width=\textwidth]{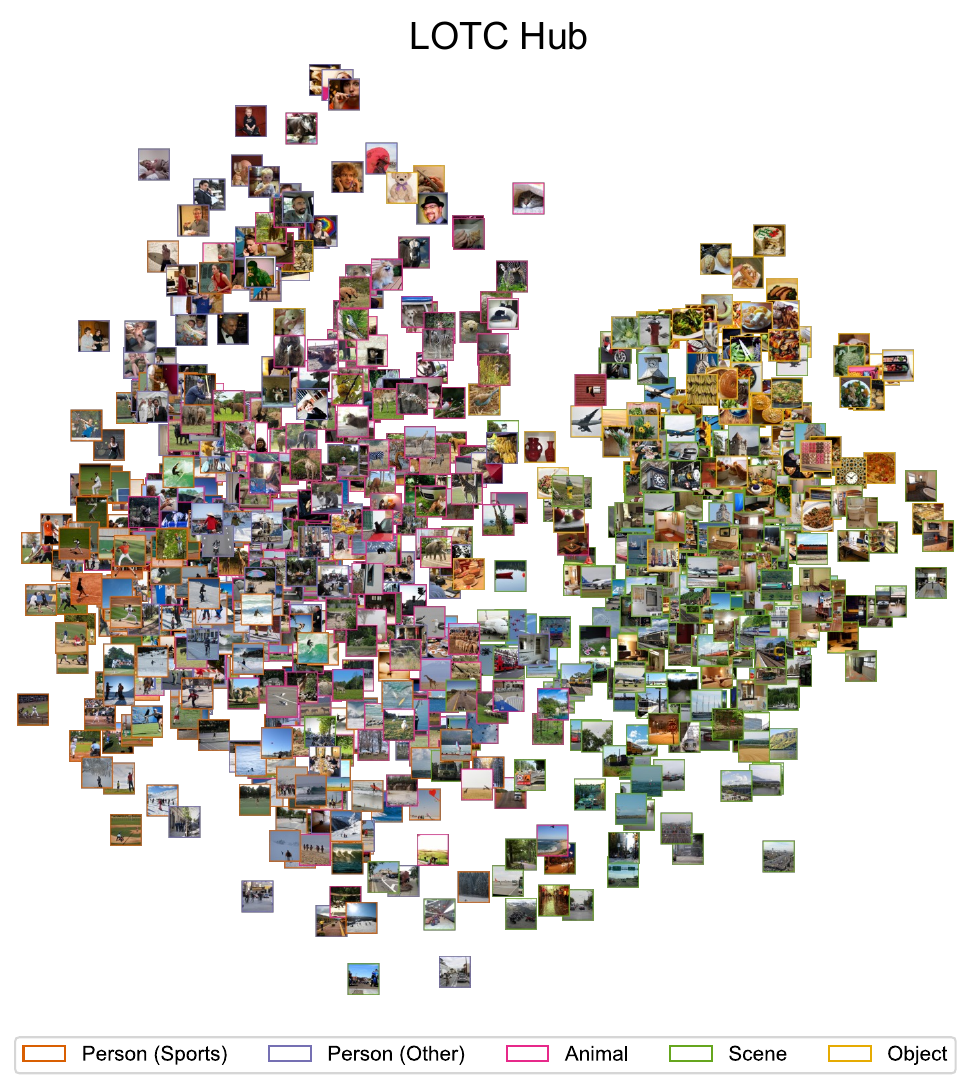}
     \caption{\textbf{Image-based atlas of the dominant representational dimensions in the LOTC Hub.}
The first two shared representational components for the LOTC hub, derived using KMCCA. Each point is an individual stimulus image, with its border colored by its general semantic category (see legend). The first KMCCA component (the x-axis) separates images containing biological agents (left cluster) from those without (right cluster). Within the biological cluster, there is a further separation, with images of animals generally located to the right of the images of people. This confirms that the LOTC hub's geometry is primarily organized around the presence and type of animate entities in a scene.
}
    \label{fig:atlas_lotc}
\end{figure}

\clearpage
\section*{\centering{Supplementary Notes}}

\begin{figure}[!b]
    \centering
    \includegraphics[width=\textwidth]{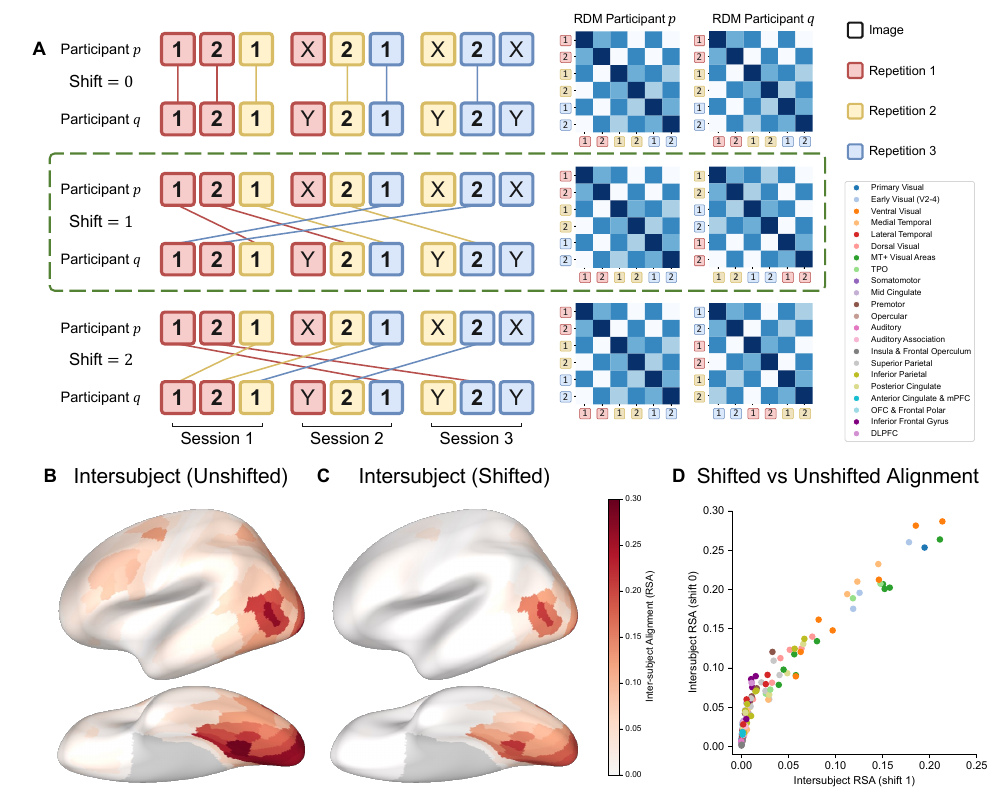}
    \caption{\textbf{Repetition-matching strategies and their impact on inter-subject RSA.}
To compute inter-subject RSA, we compared different strategies for matching the three trial repetitions for each shared image between any pair of participants.
\textbf{(A)}~Schematic of unshifted (repetition 1→1, 2→2, 3→3) versus shifted (cyclic 1→2, 2→3, 3→1) trial matching with example RDMs.
\textbf{(B, C)}Group-average cortical RSA maps under unshifted \textbf{(B)} and shifted \textbf{(C)} matching; both reveal the early-visual, ventral, and LOTC hubs.  
\textbf{(D)}~Parcel-wise comparison of shifted vs.\ unshifted RSA values shows a uniform attenuation under shifted matching, with two linear regimes reflecting stronger prefrontal attenuation.  
}
    \label{fig:extended_repetition}
\end{figure}

\subsection{Supplementary Note \thesubsection: Repetition-Matching Strategies for Inter-Subject RSA}
\label{subsection:repetition-matching}

The Natural Scenes Dataset (NSD) presents 10,000 unique images across up to 40 scanning sessions per participant (30,000 trials total). A subset of 1,000 images (3,000 trials) was shared by all eight participants and always appeared in the same trial positions--interleaved among participant‐unique stimuli--so that each shared image was viewed under identical practice, fatigue, and session‐context conditions. This locked structure raises the possibility that inter‐subject correlations might reflect non‐perceptual factors (e.g., trial timing or memory demands) rather than purely stimulus‐driven geometry.

To isolate the stimulus‐driven component, we adopted a \enquote{shifted‐repetition} matching strategy when computing inter‐subject representational similarity (IS‐RSA). Rather than pairing each repetition index $k\leftrightarrow k$ across participants (\enquote{unshifted}), we cyclically permute so that repetition $k$ in participant $p$ matches repetition $(k \bmod 3) + 1$ in participant $q$ (i.e.\ $1\rightarrow2$, $2\rightarrow3$, $3\rightarrow1$; see Fig.~\ref{fig:extended_repetition}A). Because all shared trials occupy the same session slots, this shift preserves image identity while breaking any exact repetition‐locked confounds and also allows direct comparison with within‐subject analyses (which must compare different repetitions; see Supplementary Fig.~\ref{fig:within-subject}). Due to symmetry across participant‐pair comparisons, shifts of 1 and 2 yield identical group‐average maps.

Recomputing the group‐average cortical RSA under both unshifted (Fig.~\ref{fig:extended_repetition}B) and shifted (Fig.~\ref{fig:extended_repetition}C) schemes recovers the same three hubs--early visual cortex, ventral, and LOTC. As expected, shifting is more conservative: alignment magnitudes are uniformly lower (Supplementary Fig.~\ref{fig:extended_repetition}D), yet the relative rank‐order of all parcels remains unchanged (Spearman’s \(\rho=0.96\)). Frontal parcels (e.g.\ FEF, inferior frontal gyrus) suffer proportionally larger attenuation under shifting than occipito‐temporal hubs, suggesting that these frontal regions encode task‐ or repetition‐related variance (e.g.\ decision or attentional control) that the shifted criterion minimizes. By contrast, the robust persistence of occipital and temporal alignments underscores their stable, stimulus‐driven representational geometry.

\begin{figure}[!ht]
    \centering
    \includegraphics[width=\textwidth]{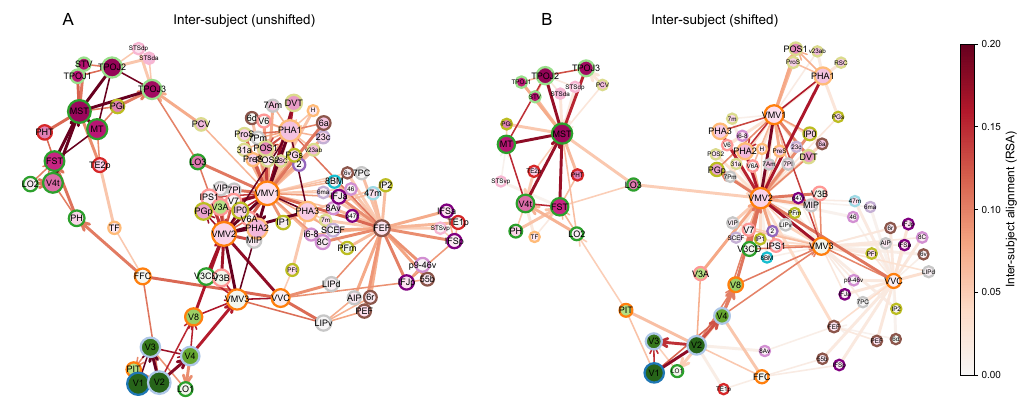}
     \caption{\textbf{Representational connectivity backbones under unshifted vs.\ shifted matching.}
    Parcels with inter‐subject RSA > 0.05 in either scheme are connected via a two‐iteration minimum‐spanning‐tree (edge cost = 1–RSA). \textbf{A}: unshifted matching; \textbf{B}: shifted matching. Both backbones preserve the Early Visual$\rightarrow$Ventral and Early Visual$\rightarrow$LOTC streams.}
    \label{fig:connectivity-shifted}
\end{figure}

To illustrate how these differences manifest in network structure, Fig.~\ref{fig:connectivity-shifted} shows representational connectivity graphs for both matching strategies. We first identified parcels whose inter-subject RSA exceeded an absolute threshold of $r=0.05$ in either the shifted (shift = 1 or 2) or unshifted maps. We then constructed a reduced connectivity matrix among this common set of nodes and extracted a two-iteration minimum‐spanning‐tree backbone (using edge weights defined as $1 - \mathrm{RSA}$; see main Methods). The core two-stream topology (Early Visual $\rightarrow$ Ventral hub and Early Visual $\rightarrow$ LOTC hub) is preserved under both schemes. However, only the unshifted graph highlights FEF as a high-centrality node, further suggesting that frontal regions carry shared task- or repetition-locked signals that are attenuated by the shifted criterion.

While these frontal effects are exploratory--given their lower RSA magnitudes and sensitivity to thresholding--they point to important future work on how task and attention signals coexist with stimulus‐driven representations.

\subsection{Supplementary Note \thesubsection: Parcel size and the reliability of IS-RSA estimates}
\label{subsection:resampling-reliability}

\textcolor{replyblue}{HCP–MMP parcels vary substantially in surface area and voxel count, raising the concern that IS-RSA estimates might be systematically less stable in smaller parcels due to reduced signal averaging. To evaluate this, we quantified the dispersion of IS-RSA estimates under two resampling schemes that isolate distinct sources of variability: changes in stimulus sampling and trial-level measurement noise.}

\begin{figure}[htb]
    \centering
    \includegraphics[width=\textwidth]{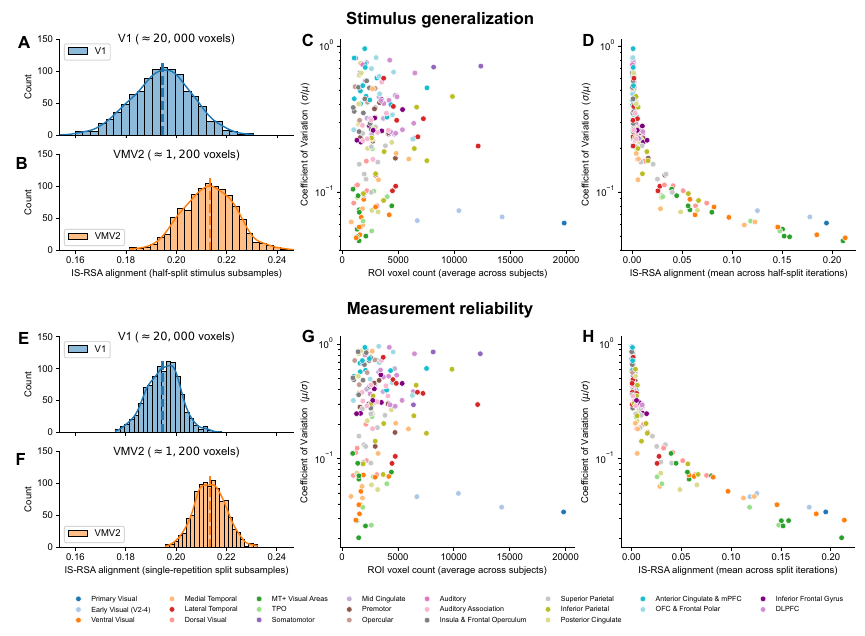}
    \caption{
    Quantifying parcel reliability via resampling. 
    \textbf{(A–D)} Stimulus generalization analysis. We computed IS-RSA on 1,000 random half-splits of the stimulus set (selecting 500 stimuli in each subsample). 
    \textbf{(A,B)} Distribution of IS-RSA scores for a representative large parcel (V1) and small parcel (VMV2). Dashed vertical lines indicate the IS-RSA estimate from the full dataset. 
    \textbf{(C)} IS-RSA coefficient of variation (CV; std/mean) plotted against parcel voxel count. 
    \textbf{(D)} CV plotted against mean IS-RSA alignment. 
    \textbf{(E–H)} Measurement reliability analysis based on randomly selecting one repetition per image across 1,000 iterations. 
    \textbf{(E,F)} Corresponding distributions for V1 and VMV2. 
    \textbf{(G)} CV versus voxel count. 
    \textbf{(H)} CV versus mean IS-RSA alignment. 
    In both analyses, the dispersion shows no dependence on voxel count but decreases as a function of the IS-RSA magnitude.
    }
    \label{fig:reliability-permutation}
\end{figure}

\textcolor{replyblue}{To assess stability with respect to stimulus sampling, we recomputed IS-RSA across 1,000 random half-splits of the image set (500 unique images per split, including all repetitions; Supplementary Fig.~\ref{fig:reliability-permutation}a-d). For each parcel, we computed the coefficient of variation (CV; standard deviation of the distribution divided by the mean) as a scale-normalized index of estimation stability.}

\textcolor{replyblue}{To isolate trial-level measurement variability while holding stimulus content fixed, we performed 1,000 iterations in which we randomly selected a single repetition per image (from the 2–3 available) and recomputed IS-RSA (Supplementary Fig.~\ref{fig:reliability-permutation}e-h). Dispersion was again summarized per parcel using the CV.}

\textcolor{replyblue}{Across both analyses, parcels with similar mean IS-RSA exhibited comparable dispersion despite large differences in voxel count (Supplementary Fig.\ref{fig:reliability-permutation}a-b, e-f). Critically, CV showed no systematic dependence on parcel size (Supplementary Fig.\ref{fig:reliability-permutation}c,g). Instead, dispersion decreased monotonically with the magnitude of IS-RSA (Supplementary Fig.~\ref{fig:reliability-permutation}d,h), indicating that parcels with stronger shared geometry yield more stable IS-RSA estimates regardless of their size. These results suggest that, under the NSD preprocessing regime, the reliability of IS-RSA estimates is primarily governed by the strength of the shared representational signal rather than by the number of voxels available for averaging.}

\subsection{Supplementary Note \thesubsection: Hemispheric Asymmetry and the Validity of Symmetric Analyses}
\label{subsection:asymmetry}

\begin{figure}[!t]
    \centering
    \includegraphics[width=\textwidth]{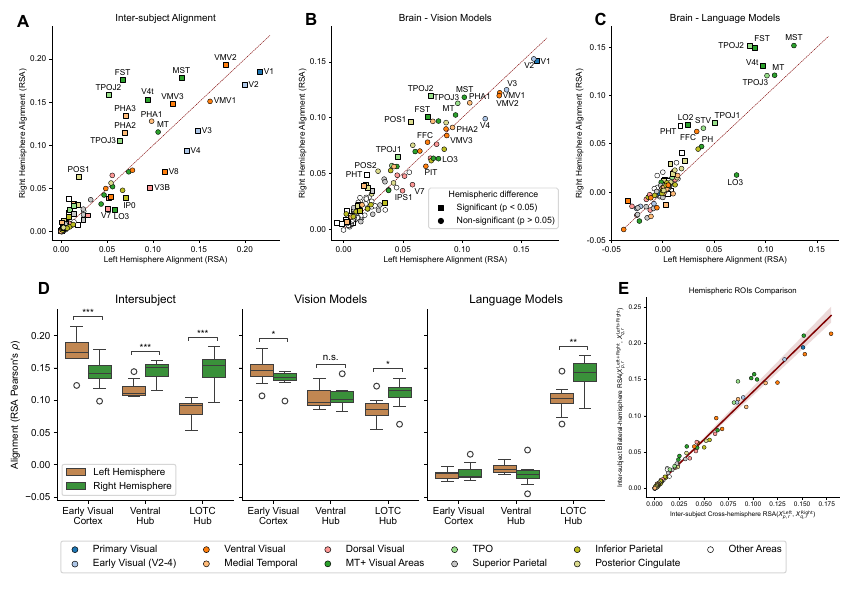}
    \caption{
\textbf{Hemispheric comparison of representational alignment.}
\textbf{(A-C)}~Comparison of representational alignment (RSA, Pearson's $r$; $N=8$, NSD) computed independently for the left (x-axis) and right (y-axis) hemispheres: \textbf{(A)} inter-subject alignment, \textbf{(B)} vision-model alignment, and \textbf{(C)} language-model alignment. Each point is a cortical parcel, colored by macro-anatomical group. The diagonal line indicates equal alignment. Squares denote parcels with a statistically significant hemispheric difference (paired $t$-test, $p < 0.05$, FDR-corrected).
\textbf{(D)}~Boxplots of alignment within principal hubs. 
\textbf{(E)} Correlation between cross-hemispheric and bilateral inter-subject alignment across parcels. Significant hemisphere effects are marked.
}
    \label{fig:hemisphere_comparison}
\end{figure}

We quantified hemispheric asymmetry in representational geometry by computing alignment measures separately for the left and right hemispheres using the HCP-MMP1.0 atlas \cite{Glasser2016}. For each cortical parcel, we assessed inter-subject alignment, vision-model alignment, and language-model alignment using the same NSD responses as in the main analyses (Fig.~\ref{fig:hemisphere_comparison}).

Across all modalities, a largely symmetric pattern was observed (Fig.~\hyperref[fig:hemisphere_comparison]{\ref*{fig:hemisphere_comparison}a-c}). The rank-order of parcels was highly consistent between hemispheres (Spearman's $\rho$: inter-subject $= 0.93$, vision $= 0.92$, language $= 0.82$). However, specific asymmetries emerged: early visual cortex (V1–V4) consistently showed stronger alignment in the left hemisphere, while the ventral and LOTC hubs (e.g., hMT+, LO, TPOJ) showed stronger alignment in the right hemisphere (Fig.~\hyperref[fig:hemisphere_comparison]{\ref*{fig:hemisphere_comparison}d}). These lateralization trends were stable across inter-subject and model-based analyses.

To determine whether these asymmetries reflected fundamental differences in representational geometry or merely differences in signal strength, we compared cross-hemispheric RSA (e.g., subject $p$’s left vs subject $q$’s right) to bilateral RSA. These measures were near-perfectly correlated across parcels ($r = 0.99$, Fig.~\hyperref[fig:hemisphere_comparison]{\ref*{fig:hemisphere_comparison}e}), with no parcel deviating substantially from the regression line. This demonstrates that combining hemispheres increases overall alignment strength (likely via noise averaging) without distorting the underlying representational geometry; no region exhibited a lateralized geometry that was absent in the contralateral hemisphere.

Together, these results validate our strategy of pooling hemispheres for the main analyses: aggregation maximizes statistical power for multivariate comparisons without introducing representational artifacts, while still allowing for targeted lateralization analyses where relevant.

\subsection{Supplementary Note \thesubsection: Power-Law Attenuation Model of RSA Metrics}
\label{section:noise-ceiling}

In the main text, we compared several RSA measures--including within-subject (WS), inter-subject (IS-1: shift=1 and IS-0: unshifted), and vision-model-brain (VM)--that probed a common cortical geometry but were subject to different levels of measurement noise. While all metrics revealed a similar cortical alignment pattern, their absolute magnitudes varied considerably. Parcel-wise scatter plots (e.g., Fig.~\ref{fig:spatial}e; Fig.~\ref{fig:extended_repetition}) showed a systematic relationship between the metrics, which were tightly related by power-law curves. Although these curves differed in scale and curvature, their consistent functional form suggested that each metric captured the same latent similarity structure while imposing a unique, noise-dependent attenuation.

\begin{figure}[b]
    \centering
    \includegraphics[width=\linewidth]{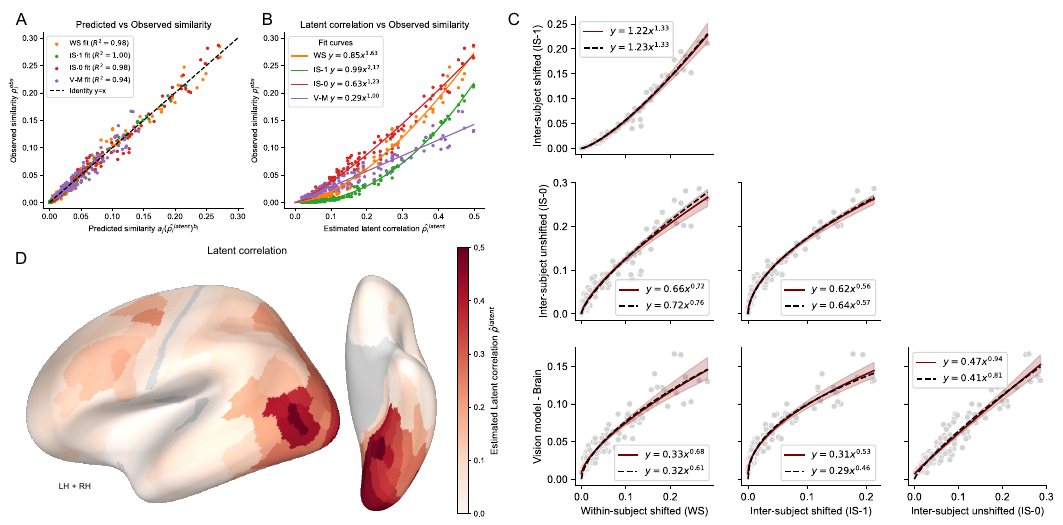}
    \caption{
    \textbf{Power-law attenuation model.}
    \textbf{(A) Goodness-of-fit.} Parcel-wise observed RSA values (y-axis) plotted against predictions from the joint power-law model (Eq.~\ref{eq:S1}; x-axis). Colors denote metrics (WS: within-subject; IS-1: inter-subject shifted; IS-0: inter-subject unshifted; VM: vision model). Legend indicates explained variance ($R^2$).
    \textbf{(B) Attenuation curves.} Estimated latent correlation ($\hat{\rho}^{\mathrm{latent}}$, x-axis) vs. observed correlation (y-axis). Colored lines show the fitted metric-specific attenuation curves ($y=a_{j}x^{b_{j}}$).
    \textbf{(C) Pairwise comparisons.} Scatter plots for six representative metric pairs (gray points). Each plot includes an independent power-law fit (solid red line, 95\% CI) and the prediction from the joint model (dashed black line, Eq.~\ref{eq:S2}), demonstrating high correspondence.
    \textbf{(D) Latent similarity map.} Cortical surface map of the estimated latent correlation values, $\hat{\rho}^{\mathrm{latent}}_{r}$ (HCP-MMP1 atlas, symmetric parcels). The model recovers the three major hubs (EVC, Ventral, LOTC) and suggests distinct signal in prefrontal regions after accounting for metric-specific attenuation.
    }
    \label{fig:noise-ceiling}
\end{figure}

We formalize this observation by modeling the correlation measured for parcel $r$ with metric $j$ as a power-law transform of a latent correlation:
\begin{equation}
\rho^\mathrm{obs}_{rj}=a_j\bigl(\rho^\mathrm{latent}_{r}\bigr)^{\,b_j} + \epsilon_{rj},
\label{eq:S1}
\end{equation}
where:
\begin{itemize}
    \item $\rho^\mathrm{latent}_{r}\in[0,1]$ is the metric-independent similarity common to all measures;
    \item $a_j\in[0,1]$ is a linear attenuation factor for metric $j$;
    \item $b_j\ge1$ captures non-linear signal compression. A value of $b_j=1$ corresponds to simple linear attenuation, whereas $b_j>1$ creates a convex curve that suppresses weak latent correlations more strongly than strong ones. This constraint is motivated by the observation that noisier metrics (e.g., IS-shift) systematically underestimate weakly aligned parcels relative to cleaner metrics (e.g., WS), consistent with a noise regime that disproportionately obscures weaker signals;
    \item $\epsilon_{rj}$ is a residual term absorbing parcel–metric–specific variability not captured by the deterministic power-law component.
\end{itemize}

We jointly fitted the parameters of Eq.~\ref{eq:S1} to four core metrics (WS, IS-0, IS-1, VM) using data from the symmetric-hemisphere ROIs. Parameter estimation was performed via the L-BFGS-B algorithm by minimizing the sum of squared residuals, with bounds: $a_j\in[0, 1]$, $b_j\in[1,\infty)$, and $\rho^\mathrm{latent}_{r}\in[0,1]$.

Figure \ref{fig:noise-ceiling}A demonstrates the model's high goodness-of-fit. In all cases, the model explained over 94\% of the variance ($R^2 > 0.94$), confirming that the power-law formulation accurately captures the relationship between observed data and the latent construct. Figure \ref{fig:noise-ceiling}B visualizes the distinct attenuation profile for each RSA strategy, illustrating how each metric compresses the estimated latent correlation.

\begin{figure}[p]
    \centering
    \includegraphics[width=\linewidth]{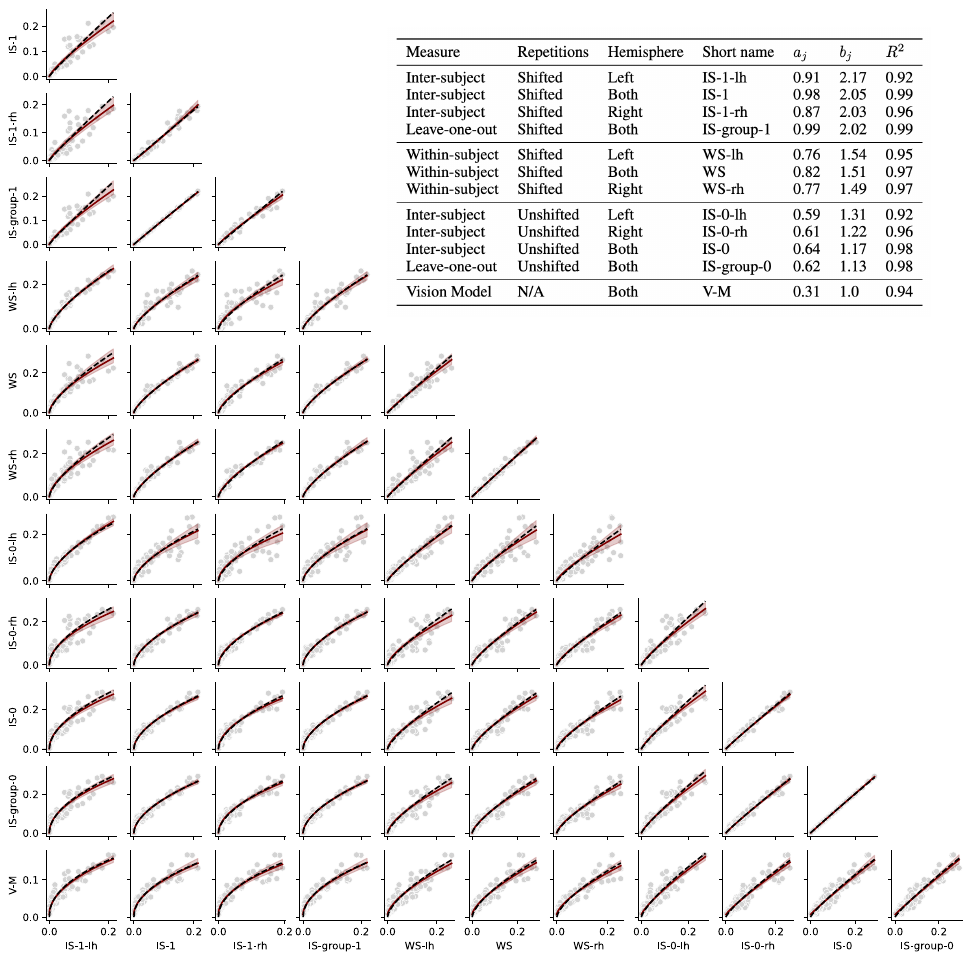}
    \caption{\textbf{Global power-law model captures pairwise relations among all RSA metrics.} 
    Scatter plots (gray points) show parcel-wise observed RSA values for all metric combinations: inter-subject (shifted and unshifted), within-subject, subject-versus-group, and vision-model correlations, each additionally split by hemisphere sampling (left, right, both). 
    Red curves depict an independent power-law fit for that specific scatter, whereas black dashed curves are the predictions from a single joint fit of Eq.~\ref{eq:S2} applied to all metrics simultaneously (no panel-specific re-fitting). 
    The accompanying table reports the fitted scale factors $a_j$, exponents $b_j$, and explained variance $R^{2}$ for every metric (model-predicted vs observed RSA). All metric trends are explained with $R^{2} > 0.92$, confirming that one latent similarity map, together with metric-specific attenuation parameters, suffices to account for the entire measurement set.}
    \label{fig:extended_ceiling}
\end{figure}

Because all metrics are modeled as functions of the same latent variable, $\rho^\mathrm{latent}_{r}$, the relationship between any two metrics, $j$ and $j'$, can be derived analytically by eliminating this common term:
\begin{equation}
y=\frac{a_{j'}}{a_{j}^{\,b_{j'}/b_{j}}}\;x^{\,b_{j'}/b_{j}},
\label{eq:S2}
\end{equation}
where $x=\rho^\mathrm{obs}_{rj}$ and $y=\rho^\mathrm{obs}_{rj'}$. Figure \ref{fig:noise-ceiling}C validates this prediction by overlaying the derived curve (Eq.~\ref{eq:S2}) on the empirical parcel scatters. The tight alignment confirms that a single latent similarity, subject to metric-specific attenuation, explains the pairwise relationships. For completeness, Figure \ref{fig:extended_ceiling} extends this validation to a set of 12 metrics, demonstrating the model's ability to unify all RSA strategies used in our analyses.

The resulting cortical map of the latent variable, $\hat{\rho}^\mathrm{latent}_{r}$, is displayed in Figure \ref{fig:noise-ceiling}D. This map represents a attenuation-corrected estimate of similarity that unifies the core RSA measures. It recovers the three-hub pattern observed in the main text, with peak correlations ($\rho^\mathrm{latent}_{r}\sim0.50$) in the EVC, Ventral stream, and LOTC. Furthermore, the model suggests that frontal areas (e.g., FEF and PFC) become more prominent after correcting for the attenuation present in raw measurements. The fact that these hubs align with those highlighted by individual metrics confirms that Eq.~\ref{eq:S1} successfully isolates a common signal while factoring out metric-specific noise.

Note that because Eq.~\ref{eq:S1} is scale-free, the absolute values of $\hat\rho^{\mathrm{latent}}$ and $a_{j}$ are defined only up to a common scaling factor. We fixed no anchor during optimization, yet repeated fits converged to the same scale, and all qualitative findings (relative parcel ranking, cross-metric relations) are invariant under rescaling.

The power-law attenuation model offers a compact, quantitative account of how measurement differences degrade RSA scores. Its success supports the view that all metrics probe a shared latent representational geometry, differing only in the fidelity with which each metric samples that geometry.

\clearpage

\subsection{Supplementary Note \thesubsection: Token‐occurrence control for language–brain RSA}
\label{subsection:tokenizer}
To quantify how much of the language–brain alignment reflects initial tokenization, we constructed RDMs from token-occurrence vectors. Each caption was encoded as a count vector---each entry recording the number of times a given token appears---and pairwise Pearson dissimilarities among these vectors yielded a tokenizer-based RDM. We applied this procedure to six vocabularies (BLOOMZ, Gemma-2, LLaMA-1/2/3; see main text Table \ref{tab:foundation_models}), covering all tokenizers used in our language models, and then computed parcel-wise RSA against NSD fMRI responses using the symmetric HCP atlas. Parcel-wise alignments were nearly identical across tokenizers (pairwise Spearman’s $\rho \ge 0.978$), so we report their average.

\begin{figure}[!ht]
    \centering
    \includegraphics[width=\linewidth]{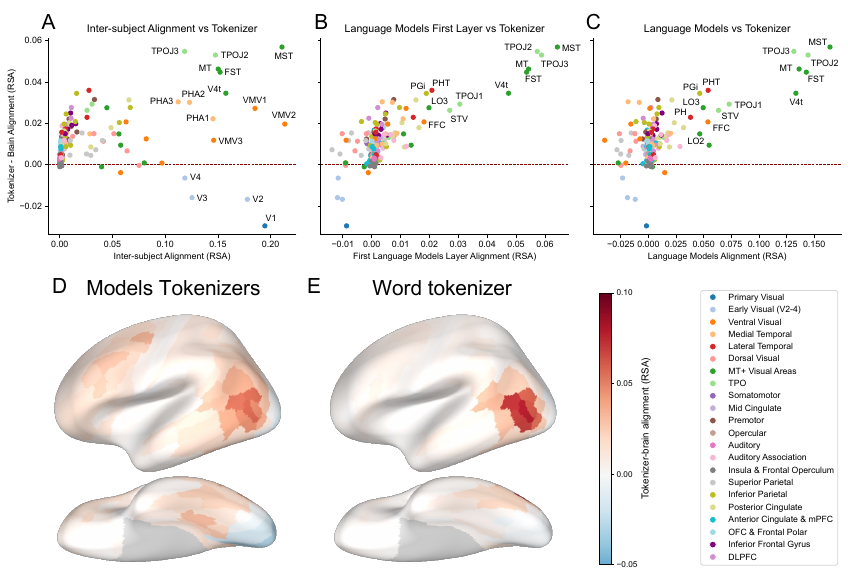}
     \caption{\textbf{Tokenizer‐level contributions to language–brain alignment.}
    We derived RDMs from token‐occurrence count vectors over each model’s vocabulary and computed parcel‐wise RSA against NSD fMRI responses (averaged across the six tokenizers used in our language models). 
    \textbf{(A)} Scatter of tokenizer‐brain vs. inter‐subject alignment, with peak values in the LOTC hub. 
    \textbf{(B)} Token‐count alignment against each model’s first‐layer alignment. 
    \textbf{(C)} Token‐count alignment against each model’s maximal alignment across layers. 
    \textbf{(D)} Cortical map of tokenizer‐brain alignment, showing pronounced LOTC localization. 
    \textbf{(E)} Alignment map for a whitespace‐based word tokenizer, which likewise peaks in LOTC, consistent with a \enquote{bag‐of‐words} explanation rather than richer compositional structure.}
    \label{fig:tokenizer_analysis}
\end{figure}

The tokenizer–brain alignment map closely matches that of full language models: it peaks in the LOTC hub and shows minimal or negative alignment in early visual parcels (Suppl. Fig.~\ref{fig:tokenizer_analysis}A,D). Token-count geometry correlates strongly with each model’s first-layer alignment (Pearson’s $\rho = 0.78$; Suppl. Fig.~\ref{fig:tokenizer_analysis}B) and with maximal alignment across layers ($\rho = 0.68$; Suppl. Fig.~\ref{fig:tokenizer_analysis}C), indicating that token frequencies account for a substantial portion of the language–brain correspondence.

Repeating the analysis with a word tokenizer---splitting captions on whitespace rather than into subword units---again yielded pronounced LOTC alignment (Suppl. Fig.~\ref{fig:tokenizer_analysis}E). These findings suggest that LOTC’s alignment to language models largely reflects sensitivity to discrete lexical tokens common across descriptions---functioning as a \enquote{bag-of-words} representation---rather than to compositional or syntactic structure \cite{yuksekgonul2023when}.

\subsection{Supplementary Note \thesubsection: Robustness to Similarity Metric (Spearman‐RSA \& CKA)}
\label{sec:metric-comparison}

\begin{figure}[!b]
    \centering
    \includegraphics[width=0.97\linewidth]{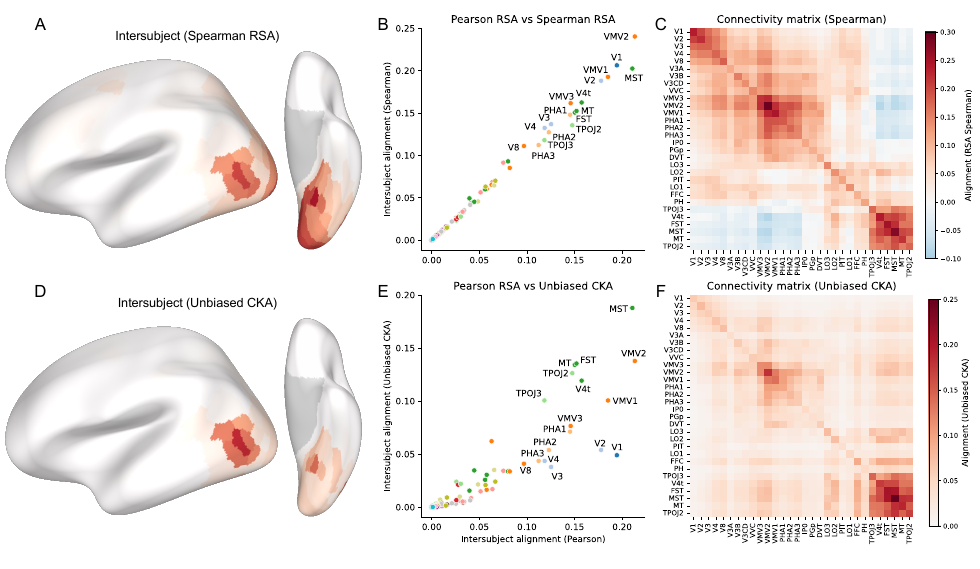}
\caption{\textbf{Comparison of alignment metrics.} \textbf{(A–C) Spearman‐RSA.} Replacing Pearson’s $r$ with Spearman’s rank correlation yields virtually identical maps: the inter‐subject alignment (A), the parcel‐wise scatter against vision‐ and language‐model alignment (B), and the representational connectivity matrix (C) all match the Pearson‐RSA findings, with parcel‐wise strengths correlating at $r=0.99$. \textbf{(D–F) Unbiased CKA.} Centered Kernel Alignment, which captures global dependence via an unbiased HSIC estimator, again highlights Early Visual, Ventral, and LOTC hubs (D). In the parcel‐wise comparison (E), CKA inflates LOTC alignment and attenuates early visual alignment--reflecting its sensitivity to low‐dimensional, dominant components--yet CKA and Pearson‐RSA remain strongly correlated ($r=0.91$). The CKA‐based connectivity matrix (F) preserves the two‐stream topology, though as a dependence measure it does not distinguish signed similarity.}
    \label{fig:alignment-comparison}
\end{figure}

Recent work has cautioned that representational similarity findings can depend not only on stimulus design but also on the choice of similarity metric---some measures probe local, pairwise geometry while others emphasize global, space-wide dependence \cite{Dujmovic2024inferring}. To test the robustness of our results to these methodological choices, we repeated key analyses using two alternatives to Pearson-RSA: (i) rank-based Spearman-RSA, which preserves only the ordering of pairwise dissimilarities, and (ii) Centered Kernel Alignment (CKA), which quantifies global statistical dependence via an unbiased Hilbert-Schmidt Independence Criterion (HSIC) estimator \cite{Song2012}.

Replacing Pearson with Spearman correlation yields a purely rank-based, nonparametric measure agnostic to absolute distance magnitudes. As shown in Supplementary Fig.~\ref{fig:alignment-comparison}A–C, the inter-subject map, the parcel-by-parcel scatter against model alignment, and the representational connectivity matrix all mirror our original Pearson-RSA results (parcel-wise Spearman's $\rho = 0.99$). This confirms that neither the linearity nor the distributional assumptions of Pearson correlation dictate our findings---our local pairwise geometry holds under a fully nonparametric test.

CKA measures dependence across entire representational spaces, prioritizing alignment along principal axes rather than isolated pairwise distances. We computed linear-kernel CKA (using the unbiased HSIC estimator \cite{Song2012}) under the same shifted-repetition protocol. Supplementary Fig.~\ref{fig:alignment-comparison}D–F shows that CKA again recovers the three hubs but inflates LOTC alignment relative to Early Visual cortex in the parcel scatter. This reflects CKA’s known sensitivity to low-dimensional, dominant components (often stronger in higher-level cortex) over the high-dimensional, finer-grained structure of early vision. Despite this spectral re-weighting, parcel-wise CKA and Pearson-RSA correlate strongly ($r=0.91$), and the CKA connectivity matrix preserves the core two-stream topology, even though CKA is strictly non-negative and thus cannot index anticorrelated geometries.

\subsection{Supplementary Note \thesubsection: Shared Component Decomposition and Partial RSA Controls}
\label{section:extended_components}

\begin{figure}[!b]
    \centering
    \includegraphics[width=\textwidth]{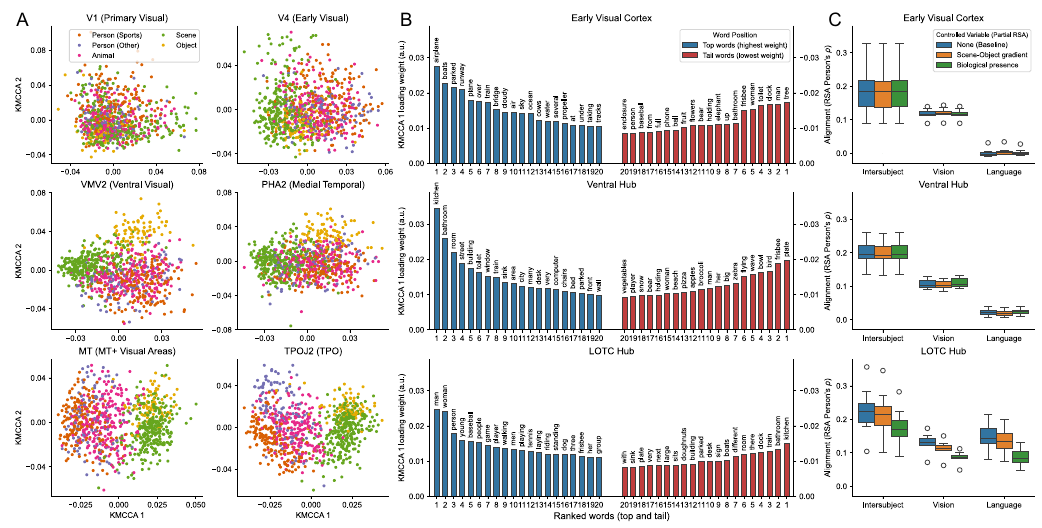}
    \caption{\textbf{Extended analysis of representational dimensions.}
    \textbf{(A)} KMCCA projections onto the top two canonical axes for six representative parcels: V1, V4, VMV2, PHA2, MT, and TPOJ2. V1 shows no clear clustering; V4 begins to separate by content; Ventral and LOTC parcels replicate the hub-level semantic gradients.
    \textbf{(B)} Words ranked by loading on the first KMCCA component (positive loadings in blue, negative in red) for each hub. Early Visual words share low-level shape features; Ventral words span scene vs.\ object terms; LOTC words distinguish animate vs.\ inanimate labels.
    \textbf{(C)} Partial RSA boxplots for the three hubs, showing inter-subject alignment under no control (baseline), controlling for the scene-object gradient, or biological presence. LOTC alignment is markedly reduced by each control, whereas Early Visual and Ventral hubs are minimally affected.}
    \label{fig:extended_components}
\end{figure}

This supplementary note details our use of Kernel Multi-view Canonical Correlation Analysis (KMCCA) \cite{Hardoon2004MKCCA} to isolate the axes driving our RSA findings. While RSA provides a single global similarity score, KMCCA decomposes this correspondence into a ranked set of orthogonal \enquote{representational channels,} revealing which latent axes contribute most strongly to the shared geometry.

We performed KMCCA using the Python package \textit{mvlearn} (v0.5.0) \cite{perry2021mvlearn} with a correlation kernel and a regularization parameter of $\lambda=0.3$, utilizing the implementation's default parameters. For each cortical hub, we formed a data matrix for each participant from the GLM-denoised single-trial $\beta$-estimates. We included only those trials for which each (image, repetition) pair was available for all eight participants. This procedure yielded a set of eight matrices ($X_{p,r}\in\mathbb{R}^{m \times v_{p,r}}$), one for each participant, all representing an identical sequence of stimulus trials. The KMCCA algorithm was then applied to this set of matrices, returning projections onto a common shared subspace. For visualization (Supplementary Fig.~\ref{fig:extended_components}A; Fig.~\ref{fig:components}A-C), we projected each subject’s data onto the first two canonical axes and then averaged the projections for the repetitions of each unique image to produce a single, stable point per stimulus.

The parcel-level KMCCA projections (Supplementary Fig.~\ref{fig:extended_components}A) confirmed that the organizational principles observed at the hub level were consistent within their constituent parcels. In Early Visual Cortex, the V1 parcel showed no clear semantic organization in its first two dimensions, whereas the V4 parcel already exhibited emergent clustering by image content. Within the Ventral hub, representative parcels such as VMV2 and PHA2 reproduced the hub-level scene-to-object gradient and showed clear organization by semantic category. Similarly, within the LOTC hub, parcels like MT and TPOJ2 showed the same primary separation between biological and non-biological stimuli. Within these main clusters, stimuli were further organized by fine-grained semantic categories, mirroring the main hub-level analysis (Fig.~\hyperref[fig:components]{\ref*{fig:components}C}).

To interpret these KMCCA axes semantically, we projected the image captions onto the shared subspace learned for each hub. We first created a vocabulary from the union of all words present in the captions ($s=11,090$ unique words), encoding each caption as a binary word-occurrence vector (a \enquote{bag-of-words} representation). We then projected this caption matrix onto the first KMCCA component derived for each hub, yielding a loading weight for each word in the vocabulary. By ranking these weights, we identified the words that contributed most positively and negatively to each dimension (Fig.~\ref{fig:extended_components}B). This word-loading analysis confirmed our interpretation of each hub's organizing principle: in Early Visual Cortex, top-weighted words were not semantically coherent but appeared to share coarse visual features; in the Ventral hub, positive tokens index scene contexts (e.g., `kitchen', `bathroom') while negative tokens name isolated objects (e.g., `plate', `frisbee'); and in the LOTC hub, tokens cleanly separated animate from inanimate labels. This analysis provides a semantic interpretation for the organization of each cluster, which can be visually confirmed in the corresponding image atlases (Supplementary Figs.~\ref{fig:atlas_evc}--\ref{fig:atlas_lotc}).

Finally, we quantified the impact of the identified dimensions on the overall inter-subject alignment using partial RSA. Denoting $\rho(X,Y)$ as the standard Pearson correlation used to compute the alignment between two RDMs, and $Z$ as a control RDM (e.g., one built from the first KMCCA component or from a binary \enquote{biological presence} vector), the partial RSA was computed as:
\begin{equation}
\mathrm{pRSA}(X,Y\mid Z)
=
\frac{\rho(X,Y)\;-\;\rho(X,Z)\,\rho(Y,Z)}
     {\sqrt{(1-\rho(X,Z)^2)\,(1-\rho(Y,Z)^2)}}.
\end{equation}
As shown in the main text (Fig.~\ref{fig:components}d-f), controlling for the continuous, data-driven KMCCA axis produced only minimal drops in the Early Visual and Ventral hubs, but a substantial reduction in the LOTC hub. A complementary analysis confirmed this result: controlling for the discrete categorical variables yielded a drop in alignment of a similar magnitude to that produced by the continuous KMCCA components (Supplementary Fig.~\ref{fig:extended_components}C).

\subsection{Supplementary Note \thesubsection: THINGS-fMRI and BOLD5000 Processing}
\label{section:comparative}

\subsubsection*{THINGS-fMRI dataset}

We processed the THINGS-fMRI dataset \cite{Hebart2023} using a pipeline analogous to that for NSD. This dataset consists of 3T fMRI recordings from three participants performing an oddball detection task on 8,740 unique object images. A subset of 100 images was repeated once per session (12 presentations total); all other images appeared only once. For our primary computations, we used only the first presentation of each of the 8,740 images.

Pre-processed, ICA-regressed single-trial estimates were obtained from the authors' public repository. We extracted voxel responses into the HCP-MMP atlas parcels provided by the THINGS team.

To control for within-session variability, inter-subject alignment was computed on a session-wise basis. For any pair of participants $(p, q)$ and parcels $(r, r')$, we computed IS-RSA by correlating the RDMs generated from the images within a given session $s$:
\begin{equation}
\mathcal{A}_{p,q,r,r',s}^{\mathrm{IS}}
=\mathrm{RSA}\bigl(X_{p,r,s},X_{q,r',s}\bigr).
\end{equation}
Group-level maps were then obtained by averaging these scores across sessions and all participant pairs. Model-brain RSA followed the same session-wise procedure. As a control, we confirmed that computing IS-RSA over the full 8,740-trial set (using $8,740 \times 8,740$ RDMs) yielded parcel-wise results that were almost perfectly correlated with our session-wise computation (Pearson's $r=0.995$). While the full-set computation produced systematically lower alignment values (linear slope $\approx 0.53$; $R^2=0.99$), this correlation validates that our session-wise approach preserves the relative spatial pattern of the findings.

\subsubsection*{BOLD5000 dataset}

We processed the BOLD5000 dataset \cite{Chang2019} using the same session-wise pipeline as for THINGS-fMRI. This dataset consists of 3T recordings from four participants performing a valence rating task on 4,916 unique images drawn from the MS-COCO, ImageNet, and SUN databases. We used the pre-processed, single-trial $\beta$-estimates provided by the authors, selecting only the first presentation of each image. Inter-subject and model-brain RSA were computed following the same session-wise procedure described above.

For voxel extraction, we registered the HCP-MMP1.0 atlas from MNI152 standard space to each subject’s native T1-weighted scan and then to their mean functional volume using FSL’s FLIRT and FNIRT. Qualitative inspection confirmed consistent parcel definitions across participants. All other preprocessing and computational steps paralleled the main NSD pipeline.

\clearpage
\subsection{Supplementary Note \thesubsection: Implementation details for large-scale RSA comparisons}
\label{section:implementation}

To efficiently handle the computational demands of large-scale RSA analyses--driven by dataset size and extensive permutation testing--we developed a GPU-accelerated implementation using PyTorch. Specifically, we reformulated representational dissimilarity matrix (RDM) calculations as optimized batch matrix operations. For each participant pair $p, q \in \mathcal{P}$ and cortical region $r \in \mathcal{R}$, we constructed stacked matrices containing flattened RDMs for all regions simultaneously, significantly reducing memory overhead and runtime compared to iterative loop-based implementations.

Given data matrices for matched trials $X_{p, r}[\boldsymbol{i}_{pq}, :]$ and $X_{q, r'}[\boldsymbol{j}_{pq}, :]$ (where the number of matched trials is $N_{\text{trials}}$), we first computed the upper-triangular vectors of dissimilarities:
\begin{equation}
\mathbf{x}^{(p)}_{pq, r} = \operatorname{vec}_u D\left( X_{p, r}[\boldsymbol{i}_{pq}, :] \right), \quad 
\mathbf{x}^{(q)}_{pq, r'} = \operatorname{vec}_u D\left( X_{q, r'}[\boldsymbol{j}_{pq}, :] \right),
\end{equation}
where $D(\cdot)$ denotes the dissimilarity function (e.g., $1 - \text{Pearson } r$) and $\operatorname{vec}_u$ denotes vectorization of the upper triangle.

Each RDM vector was centered and $l_2$-normalized before being stacked into a single matrix containing all ROIs:
\begin{equation}
\mathbf{X}^{(p)}_{pq} = 
\begin{bmatrix}
\tilde{\mathbf{x}}^{(p)}_{pq, 1} \\
\tilde{\mathbf{x}}^{(p)}_{pq, 2} \\
\vdots \\
\tilde{\mathbf{x}}^{(p)}_{pq, |\mathcal{R}|}
\end{bmatrix}
\quad \in \mathbb{R}^{|\mathcal{R}| \times \frac{N_{\text{trials}}(N_{\text{trials}}-1)}{2}},
\end{equation}
and similarly for $\mathbf{X}^{(q)}_{pq}$.

The full inter-subject connectivity matrix was then obtained efficiently via a single matrix multiplication:
\begin{equation}
\mathcal{A}^{\mathrm{IS}}_{p,q} = \mathbf{X}^{(p)}_{pq} \left( \mathbf{X}^{(q)}_{pq} \right)^\top \in \mathbb{R}^{|\mathcal{R}| \times |\mathcal{R}|}.
\end{equation}
Analogous formulations were applied for Spearman-based RSA, with rank-transformation applied to RDM vectors prior to normalization and dot products.

For brain–model alignment, we constructed similar stacked tensors for brain and model RDMs across sessions. The brain tensor was organized as:
\begin{equation}
\mathbf{X}^{(\text{brain})}_{p} \in \mathbb{R}^{|\mathcal{R}| \times S \times \frac{N_{\text{trials}}(N_{\text{trials}}-1)}{2}},
\end{equation}
and the model tensor as:
\begin{equation}
\mathbf{X}^{(\text{model})}_{m} \in \mathbb{R}^{L \times S \times \frac{N_{\text{trials}}(N_{\text{trials}}-1)}{2}},
\end{equation}
where $S$ is the number of sessions and $L$ is the number of model layers.

Brain–model RSA alignments across all regions and layers were computed via a batched contraction of the normalized RDM vectors:
\begin{equation}
\mathcal{C}_{p, m} = 
\langle \mathbf{X}^{(\text{brain})}_{p},\; \mathbf{X}^{(\text{model})}_{m} \rangle.
\end{equation}
This operation was implemented as an Einstein summation corresponding to the signature $r\,s\,d, l\,s\,d \to r\,l$ (summing over sessions $s$ and dissimilarity pairs $d$).

For permutation testing, we permuted stimulus labels according to $\sigma \in S_{N_{\text{trials}}}$ prior to RSA computation. Rather than recomputing dissimilarity matrices from raw data after each shuffle (an $O(N_{\text{trials}}^2)$ operation), we pre-calculated the equivalent permutation $\sigma'$ over the vectorized upper-triangular indices that matches the induced pairwise index shifts. This allowed us to permute the flattened RDMs directly via:
\begin{equation}
\mathbf{X}^{(p)}_{pq}[:, \sigma'],
\end{equation}
preserving the correct stimulus-pair structure with minimal computational overhead.